\documentclass[useAMS]{mn2e}
\usepackage{graphicx}	
\usepackage{amsmath}	
\usepackage{amssymb}	

\usepackage{pslatex}

\title[Project VeSElkA: HD~41076 and HD~148330]{Project VeSElkA: Abundance analysis of chemical species in HD~41076 and HD~148330\protect\thanks{Based on observations obtained at the Canada-France-Hawaii Telescope (CFHT) which is operated by the National Research Council of Canada, the Institut National des Sciences de l'Univers of the Centre National de la Recherche Scientifique of France, and the University of Hawaii. The operations at the Canada-France-Hawaii Telescope are conducted with care and respect from the summit of Maunakea which is a significant cultural and historic site.}}
\author[Khalack et al.]
       { V. Khalack$^1$\thanks{E-mail: Viktor.Khalack@umoncton.ca}, G. Gallant$^1$, C. Thibeault$^2$ \\
          $^1$D\'epartement de Physique et d'Astronomie,
               Universit\'e de Moncton, Moncton, N.-B., Canada E1A 3E9\\
          $^2$D\'epartement de Physique, Universit\'e de Montr\'eal, Montr\'eal, Qu\'ebec, Canada H3T 1J4\\
            }
\date{Accepted ???.
      Received ???;
      in original form ???}

\pubyear{2016}

\begin{document}
\label{firstpage}
\pagerange{\pageref{firstpage}--\pageref{lastpage}}
\maketitle

\begin{abstract}
A new semi-automatic approach is employed to carry out the abundance analysis of high-resolution spectra of HD~41076 and HD~148330 obtained recently with the spectropolarimetre ESPaDOnS at the CFHT. This approach allows to prepare in a semi-automatic mode the input data for the modified ZEEMAN2 code and to analyse several hundreds of line profiles in sequence during a single run. It also provides more information on abundance distribution for each chemical element at the deeper atmospheric layers. Our analysis of the Balmer profiles observed in the spectra of HD~41076 and HD~148330 has resulted in the estimates of their effective temperature, gravity, metallicity and radial velocity. The respective models of stellar atmosphere have been calculated with the code PHOENIX and used to carry out abundance analysis employing the modified ZEEMAN2 code. The analysis shows a deficit of the C, N, F, Mg, Ca, Ti, V, Cu, Y, Mo, Sm and Gd, and overabundance of Cr, Mn, Fe, Co, Ni, Sr, Zr, Ba, Ce, Nd and Dy in the stellar atmosphere of HD~41076. In the atmosphere of HD~148330, the C, N and Mo appear to be underabundant, while the Ne, Na, Al, Si, P, Ca, Ti, V, Cr, Mn, Fe, Co, Ni, Zn, Sr, Y, Zr, Ba, Ce, Pr, Nd, Sm, Eu, Gd and Dy are overabundant. We also have found signatures of vertical abundance stratification of Fe, Ti, Cr and Mn in HD~41076, and of Fe, Ti, V, Cr, Mn, Y, Zr, Ce, Nd, Sm and Gd in HD~148330.
\end{abstract}

\begin{keywords}
atomic processes -- line: formation -- line: profiles -- stars: atmospheres -- stars: chemically peculiar -- stars: individual: HD~41076 and HD~148330
\end{keywords}

\section{Introduction}

Significant portion of main sequence stars with spectral classes from B2 to F4 show abundance peculiarities of various chemical species and are commonly named chemically peculiar (CP) stars following the suggestion of George Preston \cite{Preston1974}. He has proposed to divide all known CP stars in four distinct groups depending on the presence of magnetic field in their stellar atmosphere and on the type of chemical elements that show enhanced or depleted abundance with respect to their solar abundance. Later extensive studies of CP stars revealed new types of abundance peculiarities, and a more detailed classification of CP stars was proposed by Maitzen \shortcite{Maitzen1984} and Smith \shortcite{Smith1996}. Renson \& Manfroid \shortcite{Renson+Manfroid09} have used this classification to compile an extensive list of known and suspected Ap, HgMn and Am stars.

\begin{table*}
\begin{center}
\caption{Journal of observations and measurements of the mean magnetic field. }
\label{tab1}
\begin{tabular}{lccrccc}\hline
Object & Date & HJD & $t_{exp}$ & S/N  & $<B_{\rm z}>$ &  $<N_{\rm z}>$ \\
 &  (UTC) &(2400000+)& (s)       &  Stokes I / V &   (G) &   (G) \\
\hline
HD~41076  & 2016 Feb 22 & 57440.71211 &  766 &  580/490 &  99$\pm$49 & -15$\pm$49 \\
HD~148330 & 2014 Feb 10 & 56699.09040 & 1016 & 760/500  & 114$\pm$52 &   6$\pm$52 \\
HD~148330 & 2014 Feb 15 & 56704.07140 & 1016 & 870/550  & -70$\pm$54 &   5$\pm$54 \\
\hline
\end{tabular}
\end{center}
\end{table*}

Some main sequence CP stars show abnormally strong absorption lines (due to enhanced abundance of respective chemical element) that do not vary with the phase of stellar axial rotation, while others show variability of absorption line profiles in their spectra with the period of stellar rotation. This line profile variability can be explained in terms of horizontal inhomogeneous distributions of elements abundance in stellar atmosphere \cite{Khokhlova75}. For the magnetic CP stars (BpAp stars), it was shown that the observed properties of the line profile variability can be explained in terms of overabundance and/or underabundance patches in combination with the inclined dipole model for the structure of stellar magnetic field that does not change significantly over several decades \cite{Mathys+Hubrig97,Romanyuk+14}. The significant magnetic field of BpAp stars can effectively suppress turbulence and large-scale circulation of plasma in their stellar atmosphere \cite{Alecian+Stift10,Stift+Alecian12}. This effect is even more pronounced in the slowly rotating BpAp stars. Therefore, it is commonly assumed that stellar atmospheres of BpAp stars are hydrodynamically stable. In the slowly rotating nonmagnetic CP stars that show signatures of horizontal abundance stratification the meridional circulation is assumed to be weak and they may also possess a hydrodynamically stable atmosphere \cite{Michaud+15}. This argument is not valid for the Am-Fm stars even if they rotate slowly because they possess a convective stellar atmosphere \cite{Michaud+83}. In the Am-Fm stars the observed enhanced abundance of certain chemical species is delivered to the upper atmosphere by convection from the much deeper layers (in the bottom of convection zone), where those elements are overabundant due to the effective mechanism of atomic diffusion \cite{Michaud+83,Michaud+15}.

Competition between the gravitational and radiative forces in a hydrodynamically stable atmosphere launches the atomic diffusion mechanism \cite{Michaud70} that can be responsible for the abundance peculiarities observed in CP stars. The same mechanism can cause accumulation or depletion of chemical elements at certain optical depths and lead to a vertical stratification of element abundances in the hydrodynamically stable stellar atmosphere. Presence of a strong magnetic field can intensify accumulation or depletion of chemical elements at certain optical depths and reshape a pattern of vertical stratification of element abundances \cite{Alecian+Stift10,Ryab+08,Stift+Alecian12}. Accumulation or depletion of chemical elements at certain optical depths brought about by atomic diffusion can modify the structure of stellar atmospheres. Therefore, the observed patterns of vertical abundance stratification of different chemical species are required for development and further improvement of realistic stellar atmosphere models (see, for example, Hui-Bon-Hoa et al. \shortcite{HBH+00} and LeBlanc et al. \shortcite{LeBlanc+09}) that take into account the vertical stratification of elements.

Ryabchikova et al. \shortcite{Ryabchikova+03} have shown that iron and chromium abundances increase towards the deeper atmospheric layers in $\beta$~CrB. In HD~133792 and HD~204411 the light and the iron-peak elements are concentrated in the lower atmospheric layers \cite{Ryabchikova+04}. Meanwhile the rare-earth elements (for example, Pr and Nd) are usually pushed into the upper atmosphere of Ap stars \cite{Mashonkina+05}. For the HgMn star HD~178065, Thiam et al. \shortcite{Thiam+10} have found that the manganese abundance increases towards its deeper atmospheric layers. Detection of vertical stratification of element abundances in stellar atmospheres of CP stars is an indicator of the effectiveness of the atomic diffusion mechanism \cite{Michaud70}.

To search for the signatures and to study the vertical stratification of chemical species in the atmospheres of CP stars, we have initiated Project VeSElkA \cite{Khalack+LeBlanc15a,Khalack+LeBlanc15b}, which stands for Vertical Stratification of Element Abundances.
Slowly rotating (V$\sin{i} <$ 40 km s$^{-1}$) CP stars of the upper main sequence were selected for our study using the catalogue of Ap, HgMn and Am stars of Renson \& Manfroid \shortcite{Renson+Manfroid09}.
This condition was imposed because a star with slow axial rotation has a higher probability to possess a hydrodynamically stable atmosphere where the atomic diffusion mechanism can produce vertical stratification of element abundances. Small value of V$\sin{i}$ results in a number of narrow and mostly unblended line profiles in stellar spectra that we can employ for our abundance analysis \cite{Khalack+13,Khalack+14}.

In this article we present our results from spectral analysis of HD~41076 and HD~148330 (DQ Dra) that were recently observed in the frame of the VeSElkA project. Abundance peculiarity of chemical species in stellar atmosphere of HD~148330 has been first reported by \v{Z}i\v{z}\v{n}ovsk\'{y} \shortcite{Ziga80} indicating an enhanced abundance of Ti, V, Cr, Ni and rare earth elements (REE). According to \v{Z}i\v{z}\v{n}ovsk\'{y} \& Romanyuk \shortcite{Ziga+Romanyuk90} it possess a weak longitudinal magnetic field that varies with the period P=4$^d$.2884. This period is twice larger than the period of photometric and spectral variability found for this star \cite{Ziga+Romanyuk90}.

The observations and the reduction procedure are considered in details in Section~\ref{obs}. The procedure through which we have determined the fundamental stellar parameters is discussed in Section~\ref{parameters}.
The fitting methodology is described in Section~\ref{fit}, while the results of abundance analysis are presented in Section~\ref{analysis}. A discussion follows in Section~\ref{discus}.

\section{Observations and data reduction}
\label{obs}

The high resolution (R=65000) Stokes IV spectra with S/N$>$500 were obtained recently with ESPaDOnS (Echelle SpectroPolarimetric Device for Observations of Stars) 
in the spectral domain from 3700\AA\, to 10000\AA\, employing the deep-depletion e2v device Olapa (see Table~\ref{tab1}). The optical characteristics of the spectrograph as well as the instrument performances are described by Donati et al. \shortcite{Donati+06}\footnote{For more details about this instrument, the reader is invited to visit
{\rm www.cfht.hawaii.edu/Instruments/Spectroscopy/Espadons/}}.
The obtained spectra were reduced using the dedicated software package Libre-ESpRIT \cite{Donati+97} which yields both the Stokes I spectrum and the Stokes V circular polarisation spectrum.

\begin{table*}
\begin{center}
\caption{List of fundamental stellar parameters derived from the fitting of Balmer line profiles using two grids of synthetic fluxes. One grid was simulated with the code PHOENIX16
and another one with the code PHOENIX15 (see Section~\ref{parameters}).
}
\label{tab2}
\begin{tabular}{lccrcccc}\hline
Object & $T_{\rm eff}$, & $\log(g)$ & $V \sin{i}$ & $V_{\rm r}$& [M/H] &$\chi^2/\nu$& PHOENIX\\
 & K & & km s$^{-1}$ & km s$^{-1}$ & & & version \\
\hline
HD~41076  & 9483$\pm$100 & 3.69$\pm$0.10 &  7$\pm$2$^a$ & -14.0$\pm$0.5 & -0.50$\pm$0.10 & 0.6213 & 15\\
          & 9424$\pm$100 & 3.71$\pm$0.10 &  7$\pm$2$^a$ & -14.0$\pm$0.5 & -0.50$\pm$0.10 & 0.4863 & 16\\
HD~148330 & 9303$\pm$100 & 3.70$\pm$0.10 & 10$\pm$2$^a$ & -3.3$\pm$0.5 &  0.00$\pm$0.10 & 0.6286 & 15 \\
          & 8965$\pm$100 & 3.62$\pm$0.10 & 10$\pm$2$^a$ & -3.3$\pm$0.5 &  0.00$\pm$0.10 & 0.5708 & 16 \\
\hline
\end{tabular}
\end{center}
{\it Notes:} $^a$From preliminary analysis of Si\,{\sc ii} line profiles. 
\end{table*}

\begin{figure*}
\begin{center}
\begin{tabular}{cc}
\includegraphics[width=4.6in,angle=-90]{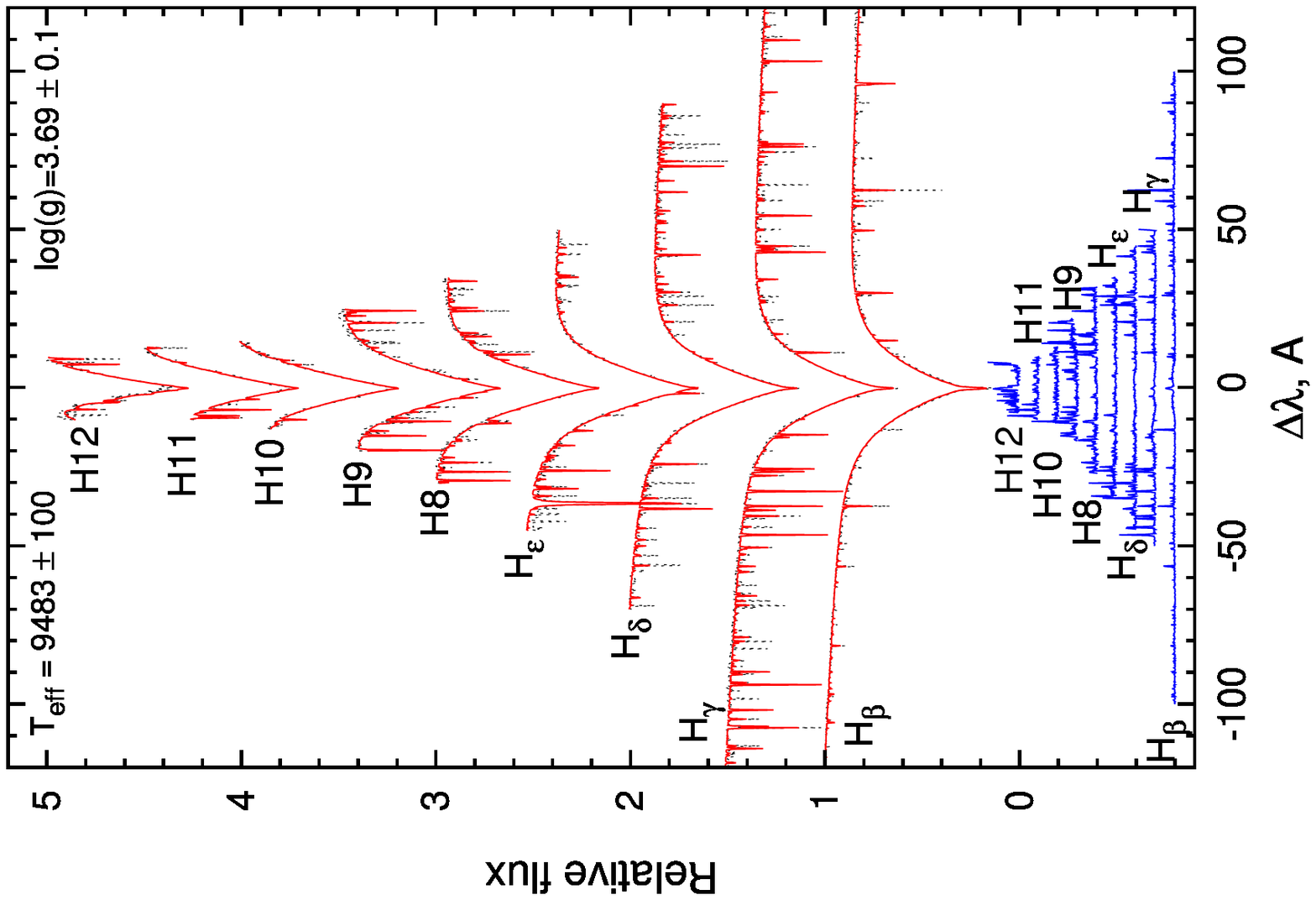} &
\includegraphics[width=4.6in,angle=-90]{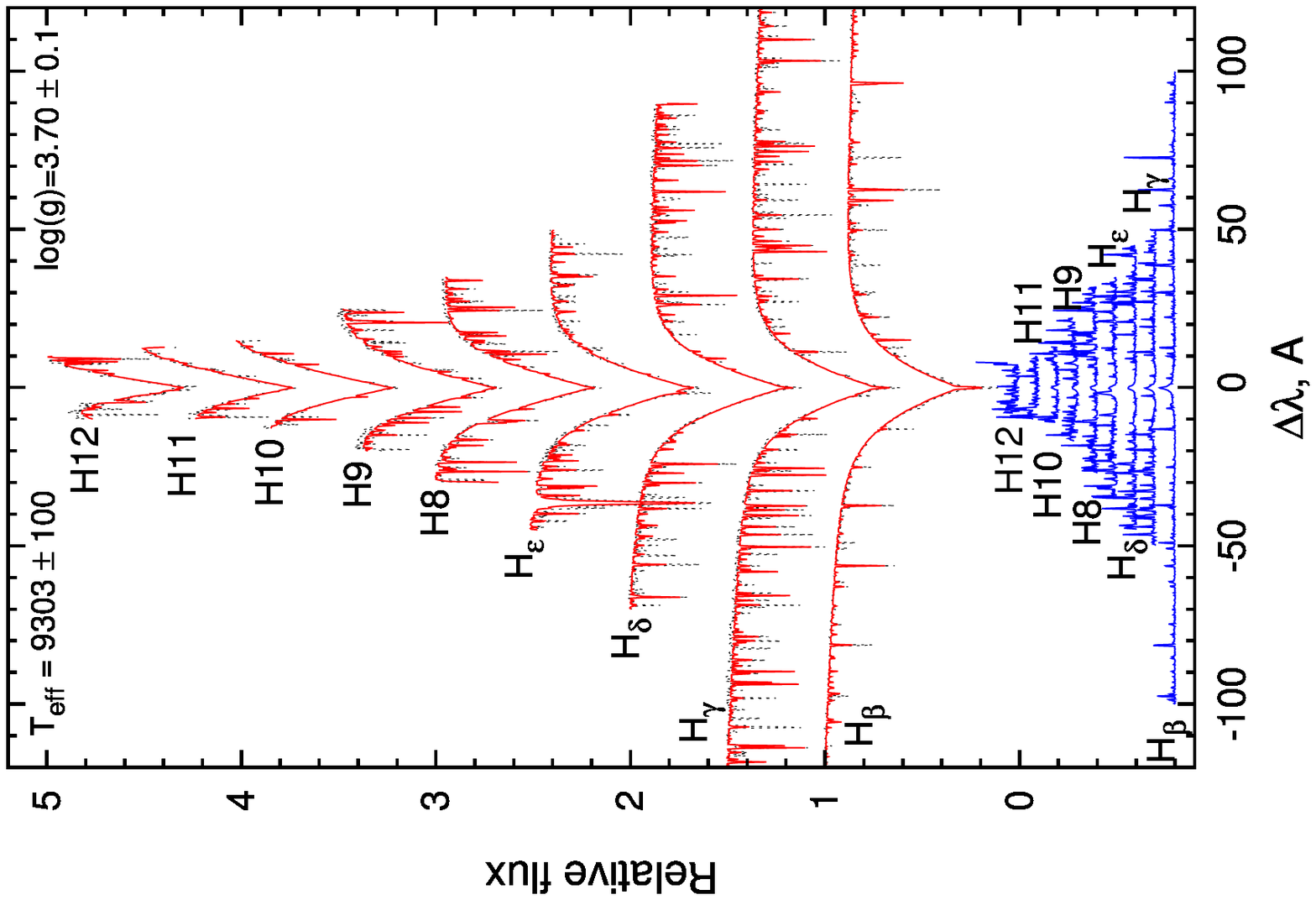} \\
\end{tabular}
\caption{ The observed Balmer line profiles (thick line) of HD~41076 (left) and HD~148330 (right) are relatively well fitted by the synthetic spectra (thin dotted line) that correspond to $T_{\rm eff}$ = 9483 K, $\log{g}$ = 3.69, [M/H]= -0.5 ($\chi^2/\nu$ = 0.6213) and $T_{\rm eff}$ = 9303 K, $\log{g}$ = 3.70, [M/H]= 0.0 ($\chi^2/\nu$ = 0.6286) respectively. Differences between the observed and synthetic spectra are shown at the bottom of each image. The Balmer line profiles are shifted by 0.5 and the differences are shifted by 0.1 for the sake of visibility. }
\label{fig1}
\end{center}
\end{figure*}

Table~\ref{tab1} provides a journal of spectropolarimetric observations HD~41076 and HD~148330, where names of the studied objects are specified in the first column. The second and the third columns present the UTC date and the HJD of data acquisition respectively. The exposure duration and the S/N ratio (in Stokes I and V spectra) are given respectively in the fourth and the fifth columns. The estimates of the mean longitudinal magnetic field and the null field in HD~41076 and HD~148330 are given in the sixth and seventh columns respectively (see Subsec.~\ref{field}).

The effective temperature and gravity were determined from the best fit of the non-normalized Balmer line profiles (see Section~\ref{parameters} and Fig.~\ref{fig1}).
In order to carry out the abundance analysis (see Section~\ref{analysis}) the reduced spectra were normalised.

\section{Estimation of fundamental stellar parameters}
\label{parameters}
\subsection{Analysis of Balmer line profiles}
\label{Balmer_par}

To determine the effective temperatures, gravities and metallicities for the studied stars (see Table~\ref{tab2}) the observed (non-normalised) Balmer line profiles were fitted with the help of FITSB2 code \cite{Napiwotzki+04} using the grids of stellar atmosphere models simulated for different $T_{\rm eff}$, $\log{g}$ and metallicity, as well as respective synthetic fluxes calculated with spectral resolution R=60000 \cite{Khalack+LeBlanc15a}. The code FITSB2 performs fitting only of the Balmer line profiles while also taking into account some strong metal lines
visible at the Balmer wings to find a model with the best $T_{\rm eff}$, $\log{g}$ and metallicity, but it does not provide an exact abundance of chemical species.
The grids of atmospheric models and synthetic fluxes have been calculated employing version 15 of the code PHOENIX \cite{Hauschildt+97}.

\begin{figure*}
\begin{center}
\begin{tabular}{cc}
\includegraphics[width=2.1in,angle=-90]{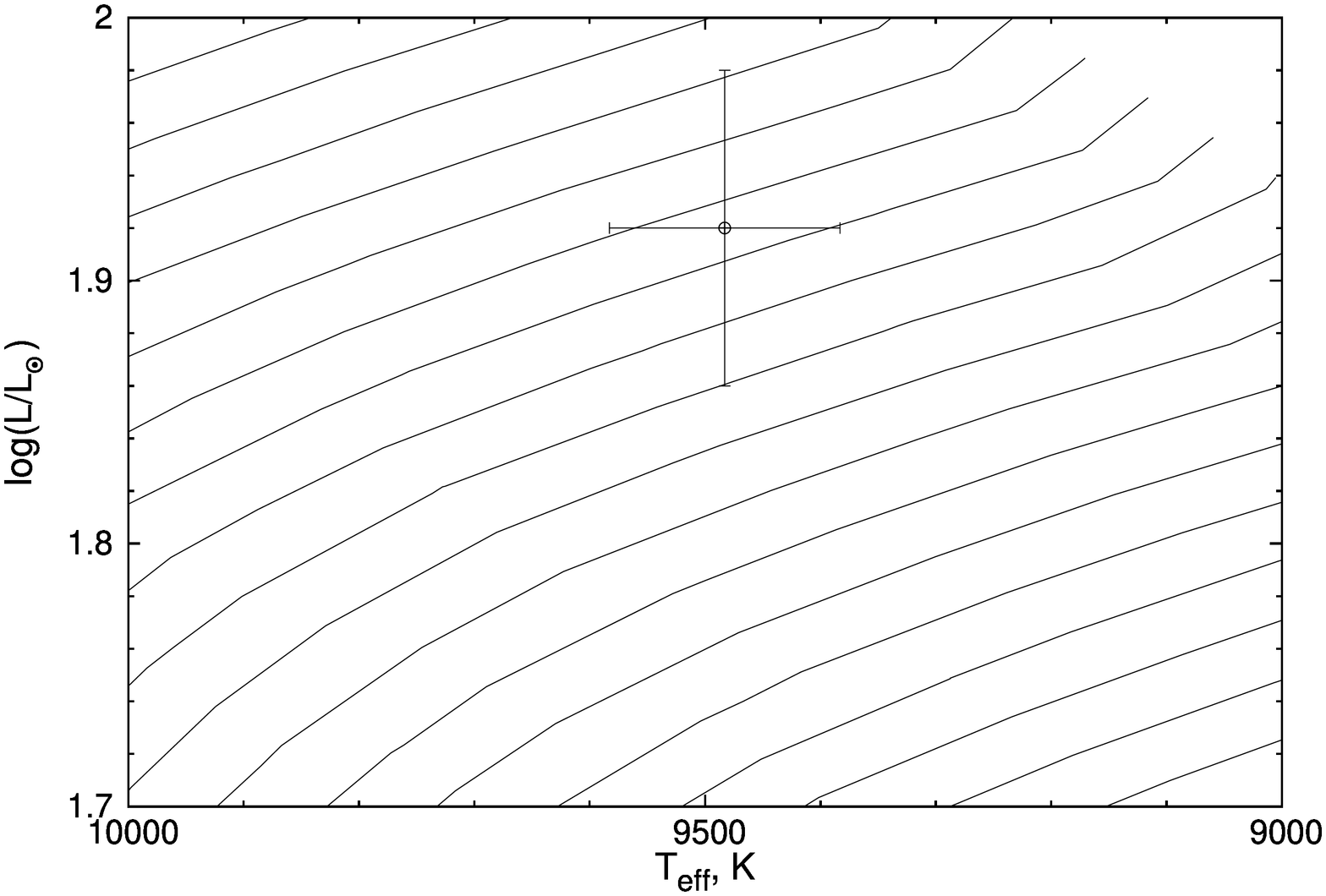} &
\includegraphics[width=2.1in,angle=-90]{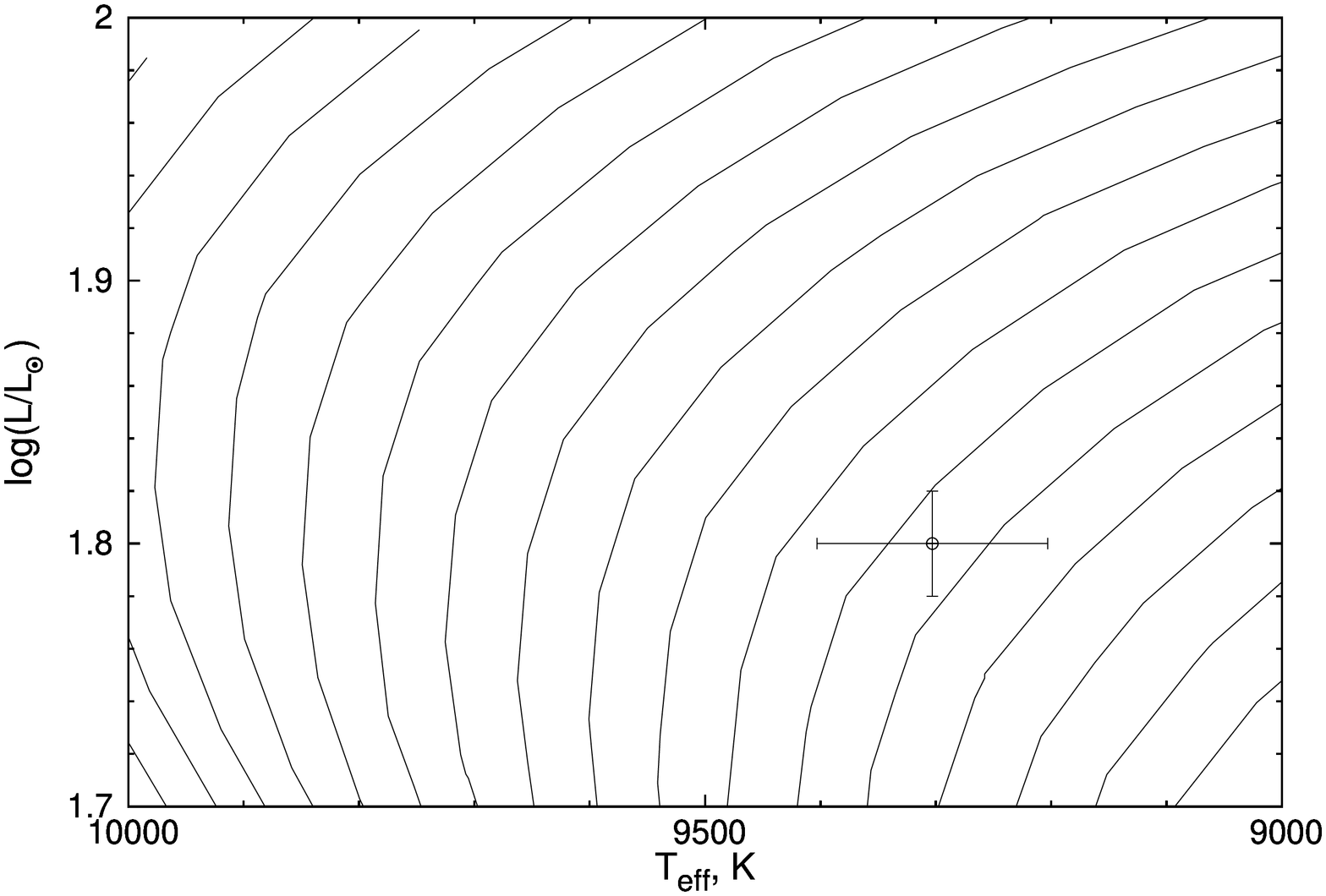} \\
\end{tabular}
\caption{  Position of HD~41076 (left) and HD~148330 (right) in the Hertzsprung-Russell diagram. The constraints for $T_{\rm eff}$ and $\log(L/L_{\odot})$ are shown by the 1$\sigma$-error bars. Isochrones are overplotted in continuous lines assuming [M/H]=-0.5 for HD~41076 and [M/H]= 0.0 for HD~148330. }
\label{fig1a}
\end{center}
\end{figure*}

The same Balmer line profiles of each star were also fitted using grids of stellar atmosphere models and respective synthetic fluxes simulated by Husser et al. \shortcite{Husser+13}.
Then we compared results obtained with two different grids of synthetic fluxes. Husser et al. \shortcite{Husser+13} have simulated a library of synthetic fluxes\footnote{Synthetic spectra are available at http://phoenix.astro.physik.uni-goettingen.de/} with spectral resolution R=500000 for stellar atmosphere models with different values of effective temperature, gravity, metallicity and abundance of the $\alpha$-elements (O, Ne, Mg, Si, S, Ar, Ca, and Ti) using the version 16 of the code PHOENIX \cite{H+B+99}. For convenience of fitting the Balmer line profiles, we have decreased the original spectral resolution of synthetic fluxes simulated by Husser et al. \shortcite{Husser+13} to R=50000, and then compiled grids of models for different metalicities and abundances of the $\alpha$-elements.

\begin{table}
\begin{center}
\caption{ Stellar parameters obtained from photometry. }
\label{tab3b}
\begin{tabular}{lcc}\hline
Parameter & HD~41076 & HD~148330 \\
\hline
Distance$^a$ (pc) & 158.98$\pm$0.07  &  119.90$\pm$0.03   \\
Age: $\log(t)$    & 8.755$\pm$0.025  &  8.705$\pm$0.015  \\
$M_{\rm V}$ (mag) & 0.03$\pm$0.15    &  0.34$\pm$0.06   \\
$\log(L_{\rm *}/L_{\rm \odot})$ & 1.92$\pm$0.06 & 1.80$\pm$0.02  \\
$T_{\rm eff}^b$ (K) & 9483$\pm$100  &  9303$\pm$100  \\
$M_{\rm *}/M_{\rm \odot}$ & 2.32$\pm$0.05 &  2.47$\pm$0.03  \\
$\log(g)$           &  3.74$\pm$0.08 & 3.86$\pm$0.06 \\
$T_{\rm eff}^c$ (K) &  9579$\pm$60   & 9292$\pm$20   \\
$T_{\rm eff}^d$ (K) &  9669$\pm$210  & 9322$\pm$180  \\
\hline
\end{tabular}
\end{center}
{\it Notes:}{  $^a$data from Hipparcos \cite{vanLeeuwen07}, $^b$results obtained from the fitting of Balmer lines using the grid of synthetic fluxes calculated with PHOENIX15 \cite{Hauschildt+97}, $^c$results obtained from the $c_{\rm 1}$ photometric temperature calibrations \cite{Napiwotzky+93}, $^d$results obtained from the $\rm (B-V)_{0}$ photometric temperature calibrations \cite{Netopil+08}.}
\end{table}

The two grids of models provide similar best fit results for the observed Balmer line profiles taking into account estimation errors for the fundamental stellar parameters except for $T_{\rm eff}$ in HD~148330 (see Table~\ref{tab2}). The fitting procedure depends very little on the value of $V\sin(i)$ used for each star. To minimise this uncertainty, we have performed a preliminary analysis of several Si\,{\sc ii} line profiles in HD~41076 and HD~148330 to obtain approximate values of $V\sin(i)$ (see Table~\ref{tab2}). These estimates of $V\sin(i)$ have been used respectively to fit the Balmer line profiles in each star. Meanwhile, the radial velocity given in Table~\ref{tab2} was determined for each star as a result of the best fit of the Balmer line profiles (see Napiwotzki et al. \shortcite{Napiwotzki+04} for details).
With the help of FITSB2 code \cite{Napiwotzki+04} we also have derived the values of $T_{\rm eff}$, $\log{g}$ and metalicity for both studied stars (see Table~\ref{tab2}).
Fig.~\ref{fig1} shows an example of the best fit of nine Balmer line profiles obtained respectively for HD~41076 and HD~148330. The differences between the observed and synthetic spectra presented at the bottom of each image show that the obtained best fit approximates the cores and the wings of Balmer lines in HD~41076 relatively well even taking into account the presence of some metals lines in the studied spectral area. In the case of HD~148330 we can see some small differences rest in the centers of several Balmer line cores, but there we also have much more metal lines that contribute to the quality of final fit increasing the value of $\chi^2/\nu$.

\subsection{Photometric temperature calibration}
\label{Photometric_T}

In the case of HD~41076, the grids of stellar atmosphere models simulated with the code PHOENIX-15 \cite{Khalack+LeBlanc15a} and with with the code PHOENIX-16 \cite{Husser+13} provide almost the same results (see Table~\ref{tab2}). Taking into account the estimation errors the derived effective temperature is close to the value $T_{\rm eff}$= 9579$\pm$60~K (see Table~\ref{tab3b}) obtained using the temperature $[c_{\rm 1}]$-calibration for Ap stars \cite{Napiwotzky+93} and the index $[c_{\rm 1}]$=1.028$\pm$0.010 determined following Str\"{o}mgren \shortcite{Stromgren66} from the $uvby\beta$ photoelectric photometry of HD~41076 \cite{Hauck+Mermilliod98}. Taking into account the parallax measured by Hipparcos \cite{ESA97,vanLeeuwen07} the distance to HD~41076 is 159~pc that results
in $\rm E(B-V)_{SFD}\approx$0.02\footnote{Service for the 3D dust mapping http://argonaut.rc.fas.harvard.edu/query} mag \cite{Schlafly+14,Green+14,Green+15} that according to Schlafly \& Finkbeiner \shortcite{Schlafly+Finkbeiner11} corresponds to $\rm E(B-V)\approx$0.018 mag.
The colour $\rm B-V\approx$-0.035 mag provided by Hipparcos for HD~41076 \cite{ESA97} is reduced to $\rm (B-V)_{0}\approx$-0.053 mag, which is very close to the limit -0.05 mag used by Netopil et al. \shortcite{Netopil+08} for photometric temperature calibration. Assuming that the interstellar reddening is slightly overestimated we have used the calibration for -0.05$\rm \leq(B-V)_{0}$ and obtained $T_{\rm eff}\approx$9669$\pm$210~K, which is again close to our result taking into account the estimation errors (see Table~\ref{tab3b}).

Using the temperature $c_{\rm 1}$-calibration for Ap stars \cite{Napiwotzky+93} and the index $[c_{\rm 1}]$=1.088$\pm$0.004 determined from the $uvby\beta$ photoelectric photometry of HD~148330 \cite{Hauck+Mermilliod98} we have estimated its $T_{\rm eff}$= 9292$\pm$20~K, which is very close to the value of effective temperature obtained employing the PHOENIX-15 grids (see Table~\ref{tab3b}). Basing on the Hipparcos parallax obtained for HD~148330 \cite{ESA97,vanLeeuwen07} we have estimated that the distance to this star is 120~pc. This estimate leads to $\rm E(B-V)_{SFD}\approx$0.005 mag \cite{Schlafly+14,Green+14,Green+15} that corresponds to $\rm E(B-V)\approx$0.0044 mag \cite{Schlafly+Finkbeiner11}. In this case, the Hipparcos $\rm B-V$ color of HD~148330 \cite{ESA97} is reduced to $\rm (B-V)_{0}\approx$-0.003 mag and the photometric temperature calibration \cite{Netopil+08} provides for this star its $T_{\rm eff}$= 9322$\pm$180~K, which also is close to our estimate. It should be pointed out that our estimate of effective temperature and the aforementioned results for $T_{\rm eff}$ obtained for this star from photometric calibrations are significantly smaller than the value $T_{\rm eff}$=9750K derived by Zabriskie \shortcite{Zabriskie77} and Cowley et al. \shortcite{Cowley+78}, but higher than the value $T_{\rm eff}$=9164K obtained by Mihalas \shortcite{Mihalas65}. Nevertheless, our estimate of effective temperature is close to the value $T_{\rm eff}$= 9400$\pm$480~K derived by \cite{Sokolov+98} from the analysis of slope of the Balmer continuum in spectrum of HD~148330.

Knowing the value of $\rm E(B-V)_{SFD}$ one can estimate an interstellar extinction \cite{Schlafly+Finkbeiner11} and for each star determine its absolute magnitude and luminosity (see Table~\ref{tab3b}) basing on the visual magnitude and distance provided by Hipparcos catalogue \cite{ESA97}. We have used the $T_{\rm eff}$ obtained from fitting of Balmer line profiles and the derived luminosity to locate the positions of the studied stars on the Hertzsprung-Russell (HR) diagram among the isochrones\footnote{For more details the reader is invited to visit {\rm stev.oapd.inaf.it/cgi-bin/cmd}.} calculated by Bressan et al. \shortcite{Bressan+12} for the metallicity [M/H]=-0.5 in the case of HD~41076 and for [M/H]= 0.0 in the case of HD~148330 (see Fig.~\ref{fig1a}).
The interpolation of data basing on the izochrones closest to the position of HD~41076 and HD~148330 in the HR diagram results in determination of their age, mass and surface gravity (see Table~\ref{tab3b}). The derived in this way values of $\log(g)$ are close to the data obtained from the fitting of Balmer line profiles taking into account the estimation errors (see Table~\ref{tab2}).

\subsection{Magnetic field estimates}
\label{field}

To estimate the longitudinal magnetic field of studied stars from their spectra-polarimetric observations (Stokes I and V spectra) we have analysed the core of $H_{\rm \alpha}$ line profile employing the method described by Landstreet et al. \shortcite{Landstreet+15}. This method has been tested using the Stokes I and V spectra of HD~16605 and HD~49299. It has provided estimates of the mean longitudinal magnetic field $<B_{\rm z}>$ in HD~16605 and HD~49299 that are similar to the data obtained by Landstreet et al. \shortcite{Landstreet+15} employing the Least Square Deconvolution method \cite{Landstreet+08}. Our estimates of the mean longitudinal magnetic field $<B_{\rm z}>$ and of the null field $<N_{\rm z}>$ for HD~41076 and HD~148330 are presented in Table~\ref{tab1}.
For both stars our measurements of $<B_{\rm z}>$ exceed the respective values of the null field, but are within the area of one-two $\sigma$ of measurement errors.

Our measurements of $<B_{\rm z}>$ in HD~148330 are in accordance with the data of \v{Z}i\v{z}\v{n}ovsk\'{y} \& Romanyuk \shortcite{Ziga+Romanyuk90} (see their Fig.~4) and Shorlin et al. \shortcite{Shorlin+02} obtained for the same star.
Nevertheless, as the obtained values of $<B_{\rm z}>$ are not statistically different from zero for both stars, we have not included magnetic field into the
procedures of line profile fitting and abundance analysis (see Section~\ref{fit}). Respectively the Zeeman and Pashen-Back effects \cite{Khalack+Landstreet12} were not taken into account during the simulation procedure.

\section{Fitting procedure}
\label{fit}

\subsection{First approach}
\label{first}

To perform abundance analysis of HD~41076 and HD~148330, we have employed two different procedures to fit the observed line profiles that belong to metals. The aim is to compare their results and to see which method is more effective.

\begin{figure*}
\begin{center}
\begin{tabular}{cc}
\includegraphics[width=2.1in,angle=-90]{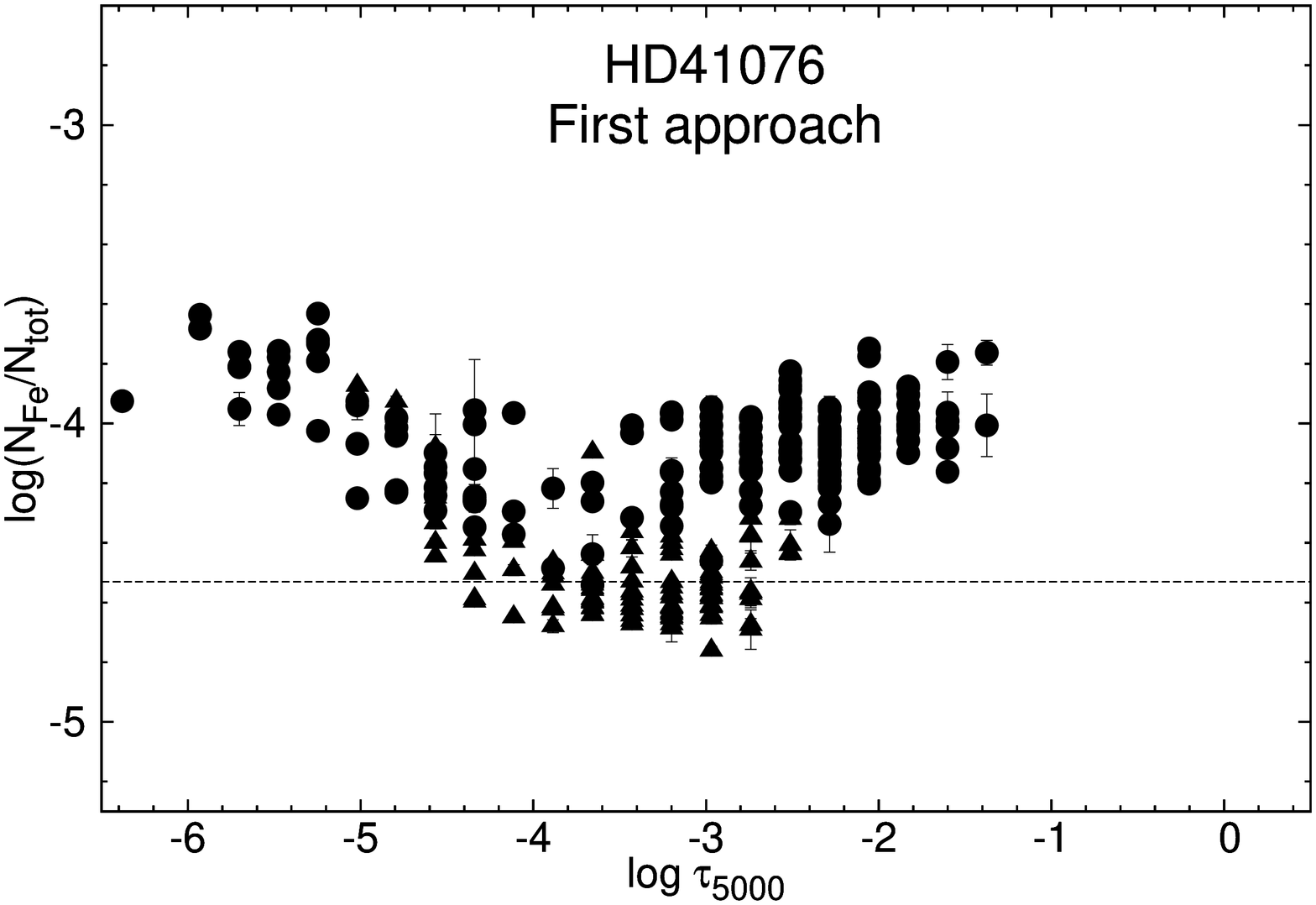} &
\includegraphics[width=2.1in,angle=-90]{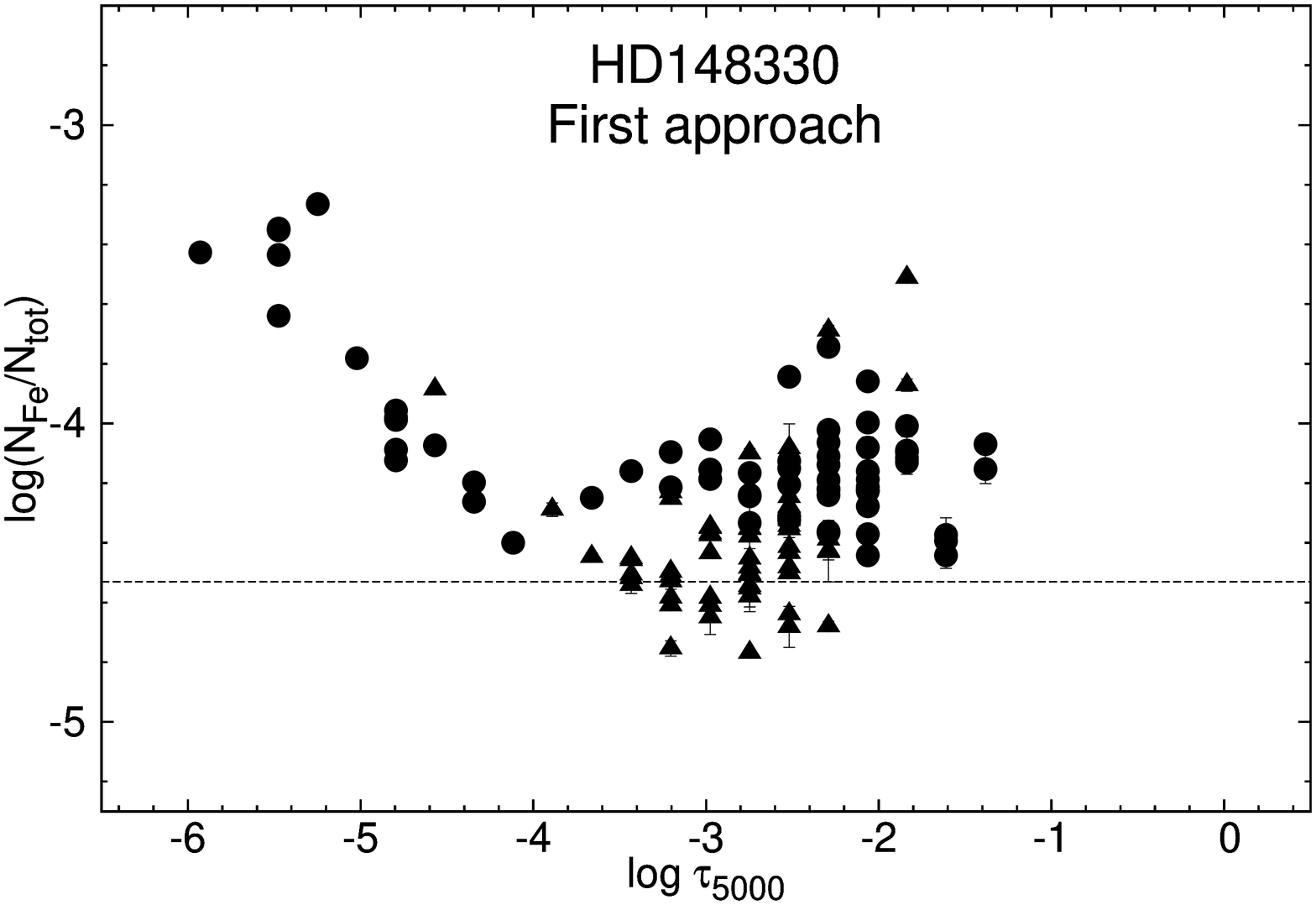} \\
\includegraphics[width=2.1in,angle=-90]{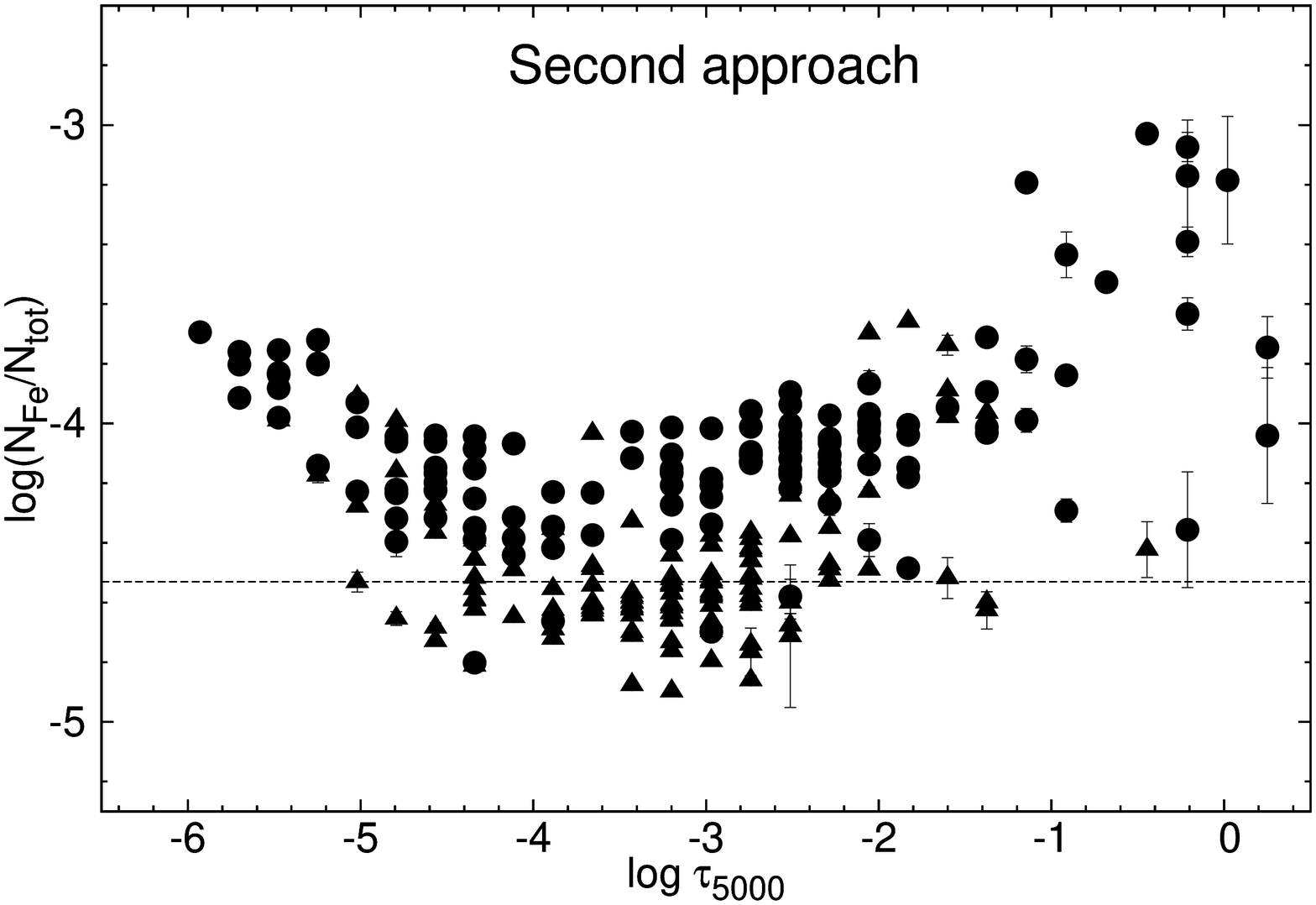} &
\includegraphics[width=2.1in,angle=-90]{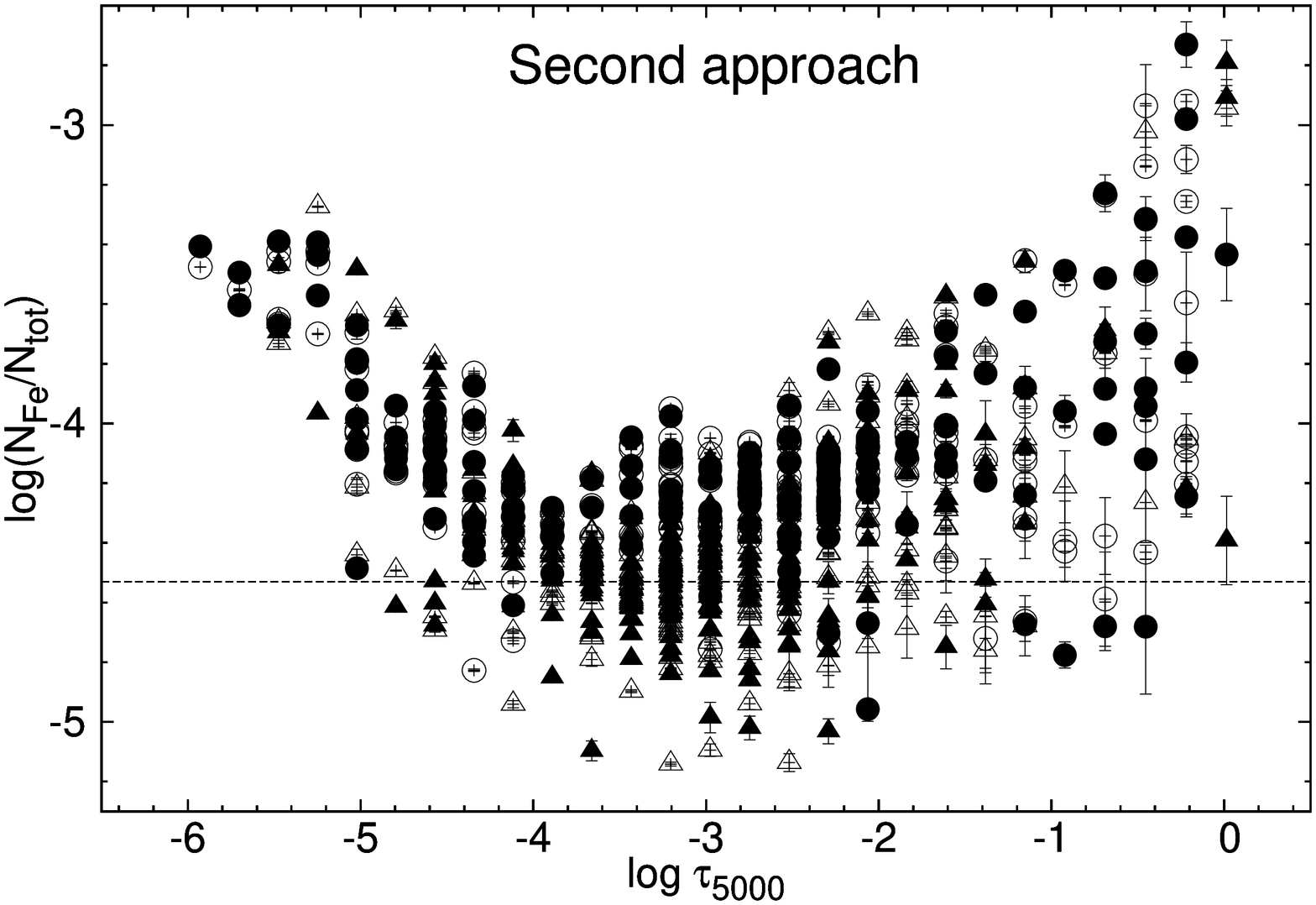} \\
\end{tabular}
\caption{ Distribution of iron abundance with optical depth in HD~41076 assuming a stellar atmosphere model with $T_{\rm eff}$= 9483 K, $\log{g}$= 3.69 (left) and in HD~148330 assuming a stellar atmosphere model with $T_{\rm eff}$= 9303 K, $\log{g}$= 3.70 (right) based on the best fit results for Fe\,{\sc i} lines (triangles) and Fe\,{\sc ii} lines (circles). Top two images show results obtained employing the first approach, while the bottom two images show results obtained with the second approach for  abundance analysis. In the case of HD~148330, results derived employing the second approach are presented by filled symbols for the first spectrum (observed on Feb.10, see Table~\ref{tab1}) and by open symbols for the second spectrum (observed on Feb.15, see Table~\ref{tab1}). Two available spectra of HD~148330 were combined while we have carried out the abundance analysis employing the first approach.}
\label{fig2}
\end{center}
\end{figure*}

\begin{table}
\centering
\caption{\label{tab3} List of spectral lines used for the abundance analysis of HD~148330 implementing the first approach (see Subsection~\ref{first}).
The full versions of this table for both studied stars are available online.}
\begin{tabular}{llllc}
\hline
Ion & $\lambda$, \AA& $\log{N_{\rm ion}/N_{\rm tot}}$& $\log gf$ & $E_{\rm l}$, $cm^{-1}$ \\
\hline
C\,{\sc i} &  4932.049 & -4.314$\pm$0.012 & -1.658 &  61981.82  \\
C\,{\sc i} &  5380.337 & -4.452$\pm$0.002 & -1.616 &  61981.82  \\
C\,{\sc i} &  9061.440 & -3.734$\pm$0.004 & -0.347 &  60352.63  \\
C\,{\sc i} &  9088.510 & -4.141$\pm$0.005 & -0.430 &  60352.64  \\
\hline
\end{tabular}
\end{table}

First approach has been previously used by Khalack et al. \shortcite{Khalack+07,Khalack+08,Khalack+10,Khalack+13,Khalack+14}, Thiam et al. \shortcite{Thiam+10} and LeBlanc et al. \shortcite{LeBlanc+15} to search for signatures of vertical stratification of element abundance in stellar atmospheres of some BHB, post-HB and CP stars (see Section~\ref{discus}). It assumes that each line profile selected for analysis is formed mainly due to a contribution of only one chemical element. Lines that belong to the different ions of the same element and contribute to a specific profile are included into simulation procedure together with lines of some other elements that appear to be significant (for example in blends). Selection procedure is usually performed manually when a researcher chooses mainly unblended (not contaminated by a contribution from other chemical species) line profiles for analysis. The fitting procedure has been performed separately for each analysed chemical element. To fit a list of selected line profiles that belong to the same element, we have employed the modified ZEEMAN2 code \cite{Khalack+Wade06,Khalack+07} that determines a radial velocity of studied star, the projection of its rotational velocity to the line of sight V$\sin{i}$ and the abundance of the chemical element independently for each profile. An example of C\,{\sc i} lines used for simulation of selected line profiles in HD~148330 with respective atomic data and derived abundances is given in Table~\ref{tab3}. The first and the second columns in the Table~\ref{tab3} specify the ion responsible for origin of analysed spectral line and its central wavelength. The 3rd, 4th and 5th columns provide respectively the found abundance of the ion, the used oscillator strength and the energy of lower term for each analysed line.

In this approach, we assume that the core of the line profile is formed mainly at line optical depth $\tau_{\rm \ell}$=1, which corresponds to a particular layer of the stellar atmosphere.
If there are several lines of the same element that contribute to the studied profile, we pick the strongest one and determine the optical depth $\tau_{\rm \ell}$=1 of its core formation.
Therefore, from the simulation of each line profile that belongs to a specific ion, we can obtain its abundance at a particular layer of the stellar atmosphere (that corresponds to a particular value of continuum optical depth $\tau_{\rm 5000}$). Taking into account that the different spectral lines (that contribute to different profiles) usually have different lower energetic levels and oscillator strengths, their cores are generally formed at different optical depths $\tau_{\rm 5000}$. In this way, we can study the vertical distribution of an element's abundance from the analysis of a large number (at least ten or more) of line profiles that belong to one or two ions of this element, which are detected in the analyzed spectrum \cite{Khalack+08}. Similar approach was used recently by \cite{Castelli+17} to study the abundance anomalies in non-magnetic Bp stars HR~6000 employing the average optical depth of line profile formation \cite{Castelli05}.

In the first approach we can use the high-resolution (R=60000) and high signal-to-noise ratio (S/N>250) spectra to derive from the fitting procedure values of the free model parameters with a precision sufficiently high to track the abundance variation with optical depth \cite{Khalack+14,Khalack+Poitras15,LeBlanc+15}. The spectra with lower resolution (R=45000) and smaller S/N ratio also can be used, but the fitting procedure will results in a lower precision of abundance estimates \cite{Khalack+07,Khalack+08,Khalack+10}. This precision will be even lower if the analysed line profile is significantly blended. Therefore, using the first approach we select for analysis mainly unblended line profiles, which we can find plenty in stars with V$\sin{i} <$ 40 km s$^{-1}$. The slowly rotating CP stars also may posses a hydrodynamically stable atmosphere where the mechanism of atomic diffusion can be effective \cite{Michaud+15}. This approach can be applied to perform abundance analysis of spectra with higher values of V$\sin{i}$ as well. Nevertheless, in this case the number of unblended line profiles will be much smaller that limits significantly our capacity to search for vertical abundance stratification of chemical species. For a star with high rotational velocity the probability to have a hydrodynamically stable atmosphere is small due to development of meridional circulation \cite{Michaud+15}.

Fig.~\ref{fig2} shows an example of distribution of iron abundance with optical depth in HD~41076 and HD~148330 obtained using this approach. A similar distribution has been derived for each analyzed chemical element. Then, taking into account this distribution, we calculated its average abundance (with respect to the solar abundance) and the estimation error (see Table~\ref{tab4}). In Table~\ref{tab4} the second and the third columns provide respectively the average abundance with the estimation errors and the number of analysed line profiles for each ion (specified in the first column) in HD~41076, while the 4th and 5th columns present the same information for HD~148330.

\begin{table}
\caption{Results of the first approach for averages abundances.} 
\label{tab4}
\begin{tabular}{l|cr|cr}\hline
Ion & \multicolumn{2}{c|}{HD~41076} & \multicolumn{2}{c|}{HD~148330}\\
  & [X/H] & N & [X/H] & N \\
\hline
C\,{\sc i}   & -0.47$\pm$0.10 &  6 & -0.59$\pm$0.31 &   4 \\
N\,{\sc i}   &                &    & -0.39$\pm$0.16 &   4 \\
O\,{\sc i}   &                &    & -0.11$\pm$0.11 &   4 \\
Mg\,{\sc i}  &                &    & -0.00$\pm$0.25 &   4 \\
Mg\,{\sc ii} & +0.16$\pm$0.25 &  4 & -0.06$\pm$0.38 &   4 \\
Al\,{\sc ii} &                &    & +0.80$\pm$0.14 &   5 \\
Si\,{\sc ii} & +0.07$\pm$0.13 &  9 & +0.52$\pm$0.23 &   7 \\
Ca\,{\sc i}  &                &    & +0.23$\pm$0.16 &  15 \\
Ca\,{\sc ii} &                &    & +0.55$\pm$0.37 &   8 \\
Ti\,{\sc ii} & -0.15$\pm$0.26 & 33 & +1.04$\pm$0.37 &  31 \\
Cr\,{\sc ii} &                &    & +0.47$\pm$0.26 &  15 \\
Fe\,{\sc i}  & +0.01$\pm$0.15 &113 & +0.14$\pm$0.24 &  58 \\
Fe\,{\sc ii} & +0.48$\pm$0.16 &190 & +0.44$\pm$0.27 &  65 \\
Ni\,{\sc i}  &                &    & +0.32$\pm$0.08 &  11 \\
Sr\,{\sc ii} &                &    & +1.13$\pm$0.08 &   2 \\
Ba\,{\sc ii} &                &    & +1.15$\pm$0.40 &   2 \\
\hline
$V \sin{i}$ (km s$^{-1}$) &   7.5$\pm$0.5 & 355 &  9.8$\pm$1.0  & 239 \\
$V_{\rm r}$ (km s$^{-1}$) & -13.8$\pm$0.5 & 355 & -3.5$\pm$0.7 & 239 \\
\hline
\end{tabular}
\end{table}

\begin{figure}
\includegraphics[width=1.65in,angle=-90]{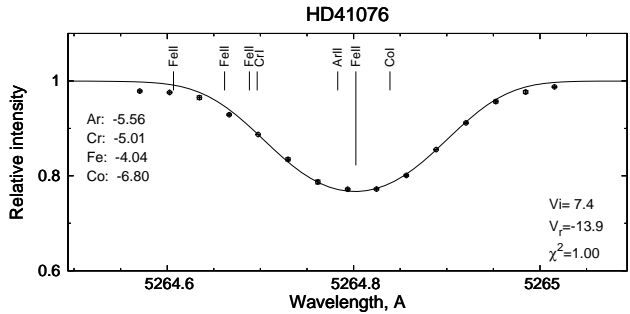}
\caption{ The observed spectrum (open circles) of HD41076 is relatively well fitted by synthetic spectrum (solid line) simulated for the stellar atmosphere model with $T_{\rm eff}$ = 9483K, $\log{g}$ = 3.69 (see Table~\ref{tab2}). Position of spectral lines that contribute to this profile is marked by vertical bars, whose length is inversely proportional to optical depths $\tau_{\rm 5000}$ of line-core formation (longer bars correspond to smaller optical depth). The best fit results for abundances of studied chemical elements are presented on the left side of the graph, while the results for radial velocity, V$\sin{i}$ and fitting quality ($\chi^2/\nu$) are given at the bottom right corner.}
\label{fig3}
\end{figure}

\subsection{Second approach}
\label{second}


Our second approach to abundance analysis uses an automatic procedure to select all the line profiles (blended or unblended with a width less than 4\AA) located in spectral regions with well visible continuum and a good S/N ratio (outside the wings of Balmer lines and outside the regions with a forest of telluric lines) for analysis. For each selected line profile, the automatic procedure uses atomic database to compile a list of lines that contribute to this profile and belongs to a set of chemical elements specified in the configuration file. Present version of automatic procedure can use up to two ionisation states of each specified element to compile a line list separately for each analysed profile. The atomic database was pre-compiled employing the NIST \cite{Kramida+15} and VALD3 \cite{Kupka+00,Ryab+15} databases.

To fit the line profiles with simulated profiles, we employed ZEEMAN2 code \cite{Landstreet88} that we have modified from the previous version \cite{Khalack+Wade06,Khalack+07} to automatically treat a list of a thousand line profiles in a consecutive mode. In this approach, each line profile is fitted taking into account lines of the first two ionisation states of 36 chemical elements (see Tables~\ref{tab5},~\ref{tab5b}) that may contribute to this profile. Therefore, we have to increase a possible number of lines (maximum 130) that can contribute to one observed profile and limit its maximal width to 4\AA. This allows us to perform complex analysis of a selected line profile and determine radial velocity of a studied star, projection of its rotational velocity to the line of sight, V$\sin{i}$, and abundance of all chemical elements whose lines contribute to this profile.

\begin{table}
\caption{ Averages abundances obtained from analysis of HD~41076 using the second approach (see Subsection~\ref{second}).}
\label{tab5}
\begin{tabular}{l|crcrcr}\hline
El. & [X\,{\sc i}/H] & N & [X\,{\sc ii}/H] & N & [X/H] & N \\
\hline
He & +0.08$\pm$0.25 &   2 &                &     & +0.08$\pm$0.25 &   2 \\ 
C  & -0.61$\pm$0.14 &   4 & -0.76$\pm$0.50 &   1 & -0.64$\pm$0.11 &   5 \\ 
N  & -0.40$\pm$0.06 &   5 & -0.74$\pm$0.11 &   2 & -0.49$\pm$0.08 &   7 \\ 
O  & +0.06$\pm$0.12 &   8 & -0.08$\pm$0.18 &   7 & -0.01$\pm$0.10 &  15 \\ 
F  & -0.81$\pm$0.19 &   3 & -1.13$\pm$0.10 &   7 & -1.04$\pm$0.10 &  10 \\ 
Ne & +0.17$\pm$0.06 &  18 &                &     & +0.17$\pm$0.06 &  18 \\ 
Na & +0.36$\pm$0.50 &   1 &                &     & +0.36$\pm$0.50 &   1 \\ 
Mg & -0.44$\pm$0.04 &   5 & -0.41$\pm$0.05 &   5 & -0.42$\pm$0.03 &  10 \\ 
Al & -0.63$\pm$0.50 &   1 & +0.20$\pm$0.50 &   1 & -0.21$\pm$0.41 &   2 \\ 
Si & -0.43$\pm$0.17 &   2 & +0.06$\pm$0.06 &  15 & +0.01$\pm$0.07 &  17 \\ 
P  &                &     & +0.30$\pm$0.34 &   6 & +0.30$\pm$0.34 &   6 \\ 
S  & +0.09$\pm$0.18 &   3 & +0.07$\pm$0.23 &   3 & +0.08$\pm$0.13 &   6 \\ 
Ar & +0.60$\pm$0.06 &   2 & -0.23$\pm$0.17 &  10 & -0.10$\pm$0.17 &  12 \\ 
Ca & -0.35$\pm$0.02 &   2 & -0.31$\pm$0.06 &   6 & -0.32$\pm$0.04 &  10 \\ 
Sc &                &     & -0.41$\pm$0.37 &   2 & -0.41$\pm$0.37 &   2 \\ 
Ti & +0.59$\pm$0.15 &   2 & -0.17$\pm$0.04 &  43 & -0.14$\pm$0.05 &  45 \\ 
V  & -1.23$\pm$0.05 &   6 & -0.99$\pm$0.32 &   4 & -1.14$\pm$0.13 &  10 \\ 
Cr & +0.89$\pm$0.13 &  28 & +0.30$\pm$0.08 &  62 & +0.49$\pm$0.07 &  90 \\ 
Mn & +0.39$\pm$0.20 &  12 & +0.08$\pm$0.18 &  15 & +0.22$\pm$0.13 &  27 \\ 
Fe & +0.03$\pm$0.02 & 134 & +0.47$\pm$0.03 & 134 & +0.25$\pm$0.02 & 268 \\
Co & +0.67$\pm$0.24 &  11 & +0.07$\pm$0.23 &   7 & +0.44$\pm$0.18 &  18 \\ 
Ni & +0.03$\pm$0.09 &  18 & +0.43$\pm$0.18 &  12 & +0.19$\pm$0.10 &  30 \\ 
Cu &                &     & -0.99$\pm$0.06 &   9 & -0.99$\pm$0.06 &   9 \\ 
Zn & +0.15$\pm$0.50 &   1 &                &     & +0.15$\pm$0.50 &   1 \\ 
Sr &                &     & +1.11$\pm$0.32 &   2 & +1.11$\pm$0.32 &   2 \\ 
Y  & -0.15$\pm$0.11 &   5 & -0.27$\pm$0.06 &   8 & -0.22$\pm$0.06 &  13 \\ 
Zr & -0.50$\pm$0.50 &   1 & +0.57$\pm$0.10 &   9 & +0.46$\pm$0.14 &  10 \\ 
Mo & -1.64$\pm$0.08 &  19 &                &     & -1.64$\pm$0.08 &  19 \\ 
Ba &                &     & +0.64$\pm$0.16 &   5 & +0.64$\pm$0.16 &   5 \\
Ce & -0.69$\pm$0.06 &   5 & +1.43$\pm$0.19 &  12 & +0.81$\pm$0.28 &  17 \\
Pr & +0.23$\pm$0.30 &   2 &                &     & +0.23$\pm$0.30 &   2 \\ 
Nd & -0.16$\pm$0.02 &   2 & +1.46$\pm$0.18 &  13 & +1.24$\pm$0.21 &  15 \\ 
Sm & -0.72$\pm$0.03 &   4 & -0.47$\pm$0.50 &   1 & -0.67$\pm$0.06 &   5 \\ 
Eu & -0.02$\pm$0.21 &   2 &                &     & -0.02$\pm$0.21 &   2 \\ 
Gd & -0.79$\pm$0.14 &   4 & -1.08$\pm$0.50 &   1 & -0.85$\pm$0.12 &   5 \\
Dy &                &     & +2.79$\pm$0.22 &   3 & +2.79$\pm$0.22 &   3 \\
\hline
\end{tabular}
\end{table}

\begin{table*}
\begin{center}
\caption{  Averages abundances obtained from analysis HD~148330 using the second approach (see Subsection~\ref{second}).}
\label{tab5b}
\begin{tabular}{l|crcrcr|crcrcr}\hline
   & \multicolumn{6}{c|}{  Feb. 10$^a$} & 
   \multicolumn{6}{c|}{  Feb. 15$^a$}\\ 
El. & [X\,{\sc i}/H] & N & [X\,{\sc ii}/H] & N & [X/H] & N & [X\,{\sc i}/H] & N & [X\,{\sc ii}/H] & N & [X/H] & N \\
\hline
He & -0.11$\pm$0.07 &   5 &                &     & -0.11$\pm$0.07 &   5 & -0.00$\pm$0.50 &   1 &                &     & -0.00$\pm$0.50 &   1 \\
C  & -0.03$\pm$0.17 &   7 &                &     & -0.03$\pm$0.17 &   7 & -0.27$\pm$0.17 &   7 &                &     & -0.27$\pm$0.17 &   7 \\
N  & -0.15$\pm$0.17 &   9 & +0.09$\pm$0.22 &   7 & -0.05$\pm$0.13 &  16 & -0.42$\pm$0.06 &   7 & -0.11$\pm$0.50 &   1 & -0.38$\pm$0.07 &   8 \\
O  & +0.03$\pm$0.12 &  14 & +0.05$\pm$0.31 &   3 & +0.03$\pm$0.11 &  17 & +0.10$\pm$0.15 &  11 & +0.02$\pm$0.19 &   5 & +0.07$\pm$0.12 &  16 \\ 
F  & +0.28$\pm$0.17 &   3 & +0.12$\pm$0.13 &  10 & +0.16$\pm$0.10 &  13 & -0.42$\pm$0.50 &   1 & +0.18$\pm$0.11 &  11 & +0.13$\pm$0.11 &  12 \\
Ne & +0.46$\pm$0.50 &   1 & +0.36$\pm$0.06 &  38 & +0.36$\pm$0.06 &  39 & -0.15$\pm$0.50 &   1 & +0.27$\pm$0.07 &  26 & +0.25$\pm$0.07 &  27 \\
Na & +0.50$\pm$0.21 &   4 &                &     & +0.50$\pm$0.21 &   4 & +0.67$\pm$0.02 &   2 &                &     & +0.67$\pm$0.02 &   2 \\
Mg & +0.18$\pm$0.11 &   7 & +0.12$\pm$0.14 &   6 & +0.15$\pm$0.08 &  13 & +0.14$\pm$0.16 &   5 & -0.08$\pm$0.07 &   6 & +0.02$\pm$0.08 &  11 \\ 
Al & +0.56$\pm$0.50 &   1 & +0.84$\pm$0.03 &   4 & +0.78$\pm$0.06 &   5 & +0.59$\pm$0.50 &   1 & +1.03$\pm$0.11 &   4 & +0.95$\pm$0.12 &   5 \\ 
Si & +0.11$\pm$0.12 &   7 & +0.49$\pm$0.07 &  18 & +0.38$\pm$0.07 &  25 & -0.12$\pm$0.12 &   6 & +0.37$\pm$0.07 &  17 & +0.24$\pm$0.07 &  23 \\
P  & +0.60$\pm$0.50 &   1 & +0.69$\pm$0.39 &   6 & +0.68$\pm$0.33 &   7 & +1.07$\pm$0.15 &   3 & +0.52$\pm$0.27 &   5 & +0.73$\pm$0.20 &   8 \\
S  & +0.30$\pm$0.08 &   8 & +0.31$\pm$0.11 &   8 & +0.31$\pm$0.06 &  16 & -0.05$\pm$0.11 &   5 & +0.24$\pm$0.15 &   7 & +0.12$\pm$0.10 &  12 \\
Ar &                &     & +0.04$\pm$0.10 &  30 & +0.04$\pm$0.10 &  30 & -0.03$\pm$0.04 &   2 & -0.04$\pm$0.11 &  27 & -0.03$\pm$0.10 &  29 \\
Ca & +0.17$\pm$0.03 &  30 & +0.20$\pm$0.13 &  10 & +0.18$\pm$0.04 &  40 & +0.22$\pm$0.05 &  25 & +0.32$\pm$0.11 &  10 & +0.25$\pm$0.05 &  35 \\ 
Sc & -0.78$\pm$0.21 &   3 & -0.04$\pm$0.09 &  11 & -0.20$\pm$0.12 &  14 & -0.40$\pm$0.30 &   5 & -0.05$\pm$0.12 &  11 & -0.16$\pm$0.12 &  15 \\
Ti & +0.93$\pm$0.13 &  22 & +0.84$\pm$0.07 &  90 & +0.86$\pm$0.06 & 112 & +1.22$\pm$0.15 &  20 & +0.75$\pm$0.05 &  81 & +0.84$\pm$0.05 & 101 \\ 
V  & +0.25$\pm$0.15 &  12 & +0.60$\pm$0.07 &  33 & +0.51$\pm$0.07 &  45 & +0.10$\pm$0.31 &   7 & +0.62$\pm$0.07 &  32 & +0.53$\pm$0.09 &  39 \\ 
Cr & +0.85$\pm$0.10 &  47 & +0.49$\pm$0.05 &  90 & +0.61$\pm$0.05 & 137 & +0.70$\pm$0.09 &  58 & +0.42$\pm$0.05 &  89 & +0.53$\pm$0.05 & 147 \\
Mn & +0.55$\pm$0.16 &  27 & +0.46$\pm$0.13 &  27 & +0.50$\pm$0.10 &  54 & +0.38$\pm$0.13 &  26 & +0.31$\pm$0.10 &  32 & +0.34$\pm$0.08 &  58 \\ 
Fe & +0.13$\pm$0.02 & 222 & +0.43$\pm$0.03 & 173 & +0.26$\pm$0.02 & 395 & +0.11$\pm$0.02 & 232 & +0.44$\pm$0.03 & 163 & +0.25$\pm$0.02 & 395 \\
Co & +0.99$\pm$0.20 &  19 & +1.30$\pm$0.24 &   9 & +1.09$\pm$0.16 &  28 & +1.35$\pm$0.17 &  21 & +1.25$\pm$0.20 &  11 & +1.31$\pm$0.13 &  32 \\ 
Ni & +0.54$\pm$0.07 &  50 & +1.03$\pm$0.08 &  24 & +0.70$\pm$0.06 &  74 & +0.57$\pm$0.06 &  58 & +1.20$\pm$0.10 &  28 & +0.77$\pm$0.06 &  86 \\ 
Cu &                &     & -0.09$\pm$0.10 &  18 & -0.09$\pm$0.10 &  18 &                &     & -0.18$\pm$0.12 &  14 & -0.18$\pm$0.12 &  14 \\
Zn & +0.66$\pm$0.02 &   2 &                &     & +0.66$\pm$0.02 &   2 &                &     & +0.61$\pm$0.03 &   2 & +0.61$\pm$0.03 &   2 \\
Sr & +2.70$\pm$0.50 &   1 & +1.56$\pm$0.32 &   2 & +1.84$\pm$0.37 &   3 &                &     & +1.14$\pm$0.14 &   3 & +1.14$\pm$0.14 &   3 \\
Y  & +0.94$\pm$0.02 &   2 & +0.98$\pm$0.07 &  22 & +0.98$\pm$0.06 &  24 & +1.22$\pm$0.27 &   3 & +0.99$\pm$0.07 &  21 & +1.02$\pm$0.08 &  25 \\
Zr & +2.40$\pm$0.08 &   2 & +1.12$\pm$0.05 &  41 & +1.18$\pm$0.06 &  43 & -0.25$\pm$0.50 &   1 & +1.17$\pm$0.05 &  39 & +1.13$\pm$0.06 &  40 \\
Mo & -0.24$\pm$0.06 &   9 &                &     & -0.24$\pm$0.06 &   9 & -0.22$\pm$0.08 &  10 &                &     & -0.22$\pm$0.08 &  10 \\
Ba &                &     & +1.50$\pm$0.23 &   7 & +1.50$\pm$0.23 &   7 &                &     & +1.51$\pm$0.18 &   4 & +1.51$\pm$0.18 &   4 \\ 
Ce & +0.27$\pm$0.10 &  10 & +1.56$\pm$0.09 &  58 & +1.37$\pm$0.10 &  68 & -0.42$\pm$0.20 &   4 & +1.26$\pm$0.09 &  45 & +1.12$\pm$0.11 &  49 \\ 
Pr & +0.39$\pm$0.33 &   2 & +1.47$\pm$0.27 &   7 & +1.23$\pm$0.27 &   9 &                &     & +2.08$\pm$0.19 &  13 & +2.08$\pm$0.19 &  13 \\
Nd & +0.33$\pm$0.09 &   2 & +1.41$\pm$0.12 &  33 & +1.35$\pm$0.12 &  35 & +0.10$\pm$0.50 &   1 & +1.52$\pm$0.10 &  33 & +1.47$\pm$0.10 &  34 \\ 
Sm & +1.06$\pm$0.21 &   5 & +2.09$\pm$0.17 &  14 & +1.82$\pm$0.17 &  19 & +0.28$\pm$0.09 &   2 & +1.80$\pm$0.15 &  20 & +1.66$\pm$0.17 &  22 \\
Eu & +0.24$\pm$0.31 &   2 & +0.93$\pm$0.28 &   4 & +0.70$\pm$0.24 &   6 &                &     & +1.22$\pm$0.27 &   6 & +1.22$\pm$0.27 &   6 \\
Gd & +0.37$\pm$0.15 &   4 & +1.16$\pm$0.10 &  14 & +0.98$\pm$0.12 &  18 &                &     & +1.26$\pm$0.10 &  24 & +1.26$\pm$0.10 &  24 \\
Dy &                &     & +0.88$\pm$0.24 &   4 & +0.88$\pm$0.24 &   4 &                &     & +0.99$\pm$0.15 &   4 & +0.99$\pm$0.15 &   4 \\
\hline
\end{tabular}
\end{center}
{\it Notes:} $^a$results for two separate spectra (see Table~\ref{tab1})
\end{table*}

There might be several lines that belong to the same element and contribute to an analyzed profile. Among these lines, we search for optical depth $\tau_{\rm \ell}$=1 that corresponds to core formation of the strongest line (in the spectral area of studied profile) of this element (taking into account its two first ionisation states) and find a corresponding value of optical depth $\tau_{\rm 5000}$ (corresponding layer of the stellar atmosphere). Abundance of that particular element derived from the best fit procedure is linked to the aforementioned optical depth $\tau_{\rm 5000}$. Therefore we can create a plot abundance versus optical depth similar to the those shown at Fig.~\ref{fig2}.
The same procedure is employed for other elements whose lines contribute to the analyzed profile. In this way we connect the derived abundance of each element to the optical depth of core formation of its strongest line in the spectral area of analysed profile.

\begin{figure*}
\begin{center}
\begin{tabular}{cc}
\includegraphics[width=2.1in,angle=-90]{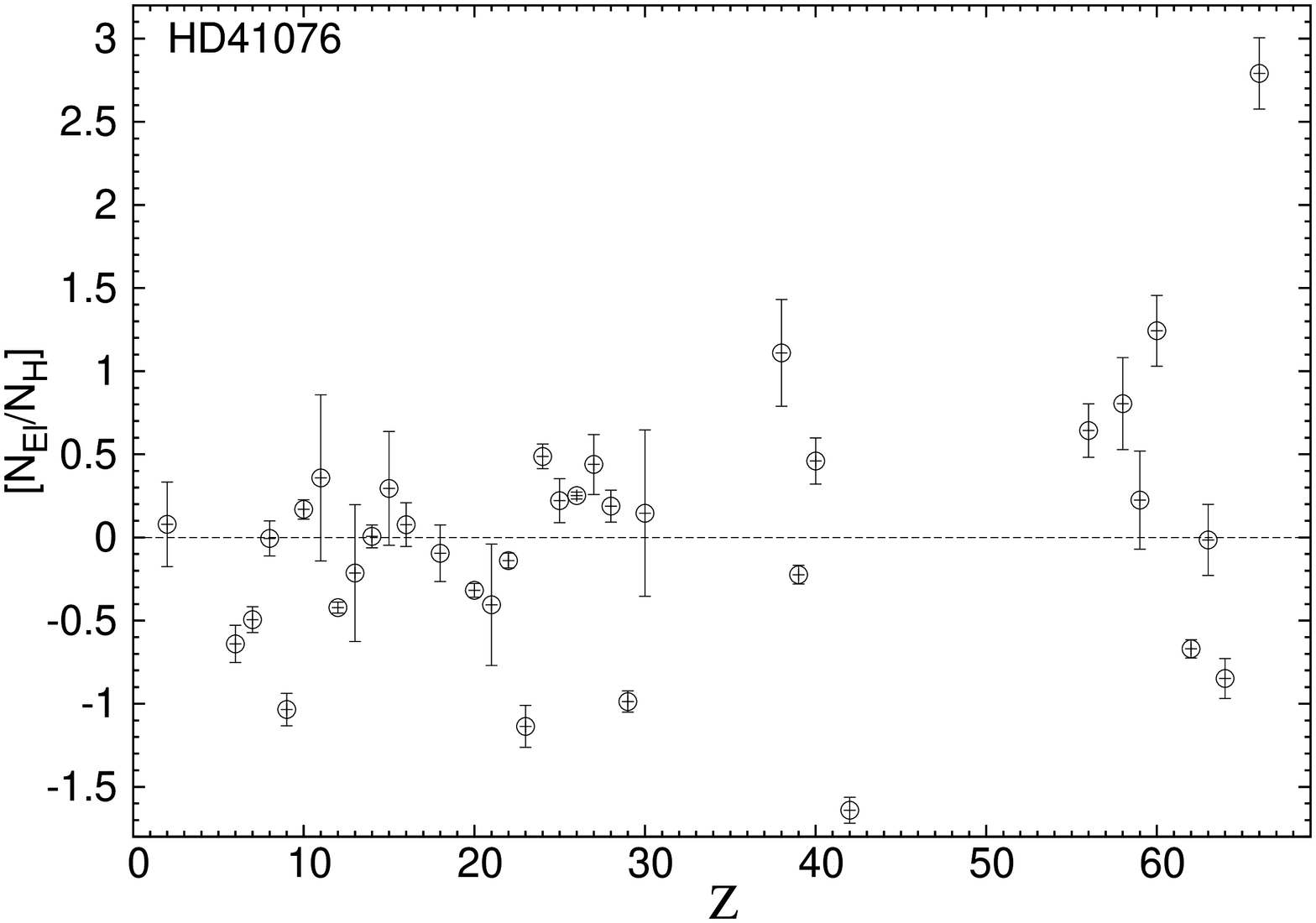} &
\includegraphics[width=2.1in,angle=-90]{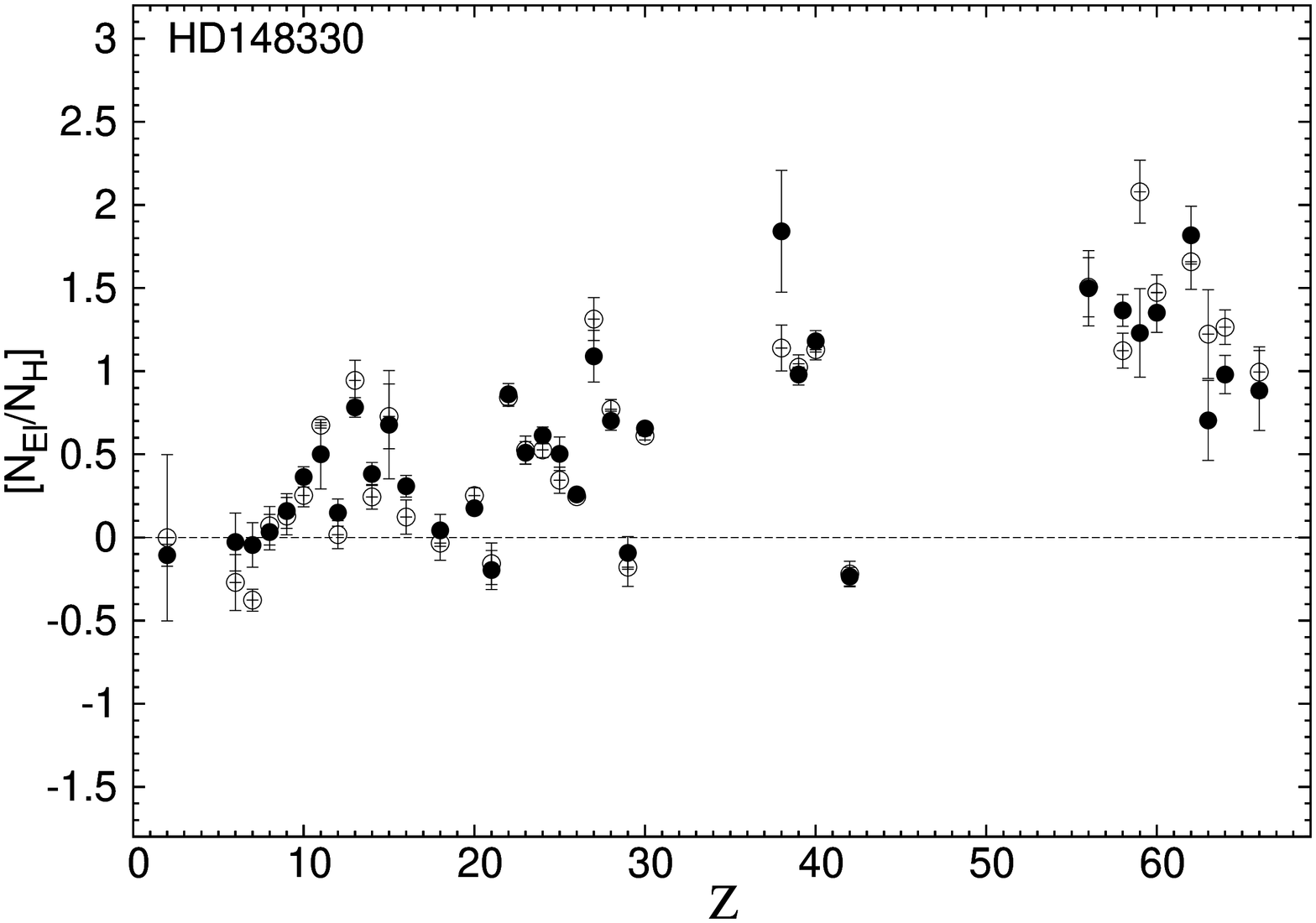}\\
\end{tabular}
\caption{  Average abundance of chemical species in HD~41076 (left panel) and HD~148330 (right panel) with respect to their solar abundance (dashed line). In the case of HD~148330, the results derived from analysis of the first spectrum (observed on Feb.10, see Table~\ref{tab1}) are presented by filled symbols and those derived from analysis of the second spectrum (observed on Feb.15, see Table~\ref{tab1}) are shown by open symbols.}
\label{fig4b}
\end{center}
\end{figure*}

Usually the studied line profiles contain one or several strong lines and some weak blends. Fig.~\ref{fig3} shows an example of the best fit of analysed profile in the area 5264-5265\AA\AA\, of HD~41076 spectrum, where abundances of studied chemical elements are presented on the left side of the graph while the values for velocities and the fitting quality ($\chi^2/\nu$) are given at the bottom right corner. This profile includes relatively strong Fe\,{\sc ii} line and several blends of ions that are well fitted using our second approach. This example shows an advantage of the second approach that allows us to properly fit even a complex line profile in an automatic regime.

The approach described above employs lines of up to 36 chemical elements to fit a studied line profile and therefore is more flexible than the first approach (see Subsection~\ref{first}). Our second approach provides abundances of several elements from the analysis of only one profile. Precision of abundance determination for elements that are responsible for weak blend is not high, but in combination with the results obtained for the same element from the best fit of other line profiles, this contributes to general picture of abundance variation with optical depth for this particular element. From our experience, the high resolution spectra obtained with ESPaDOnS (see Section~\ref{obs}) always contain more spectral measurements (data at certain wavelengths) in a studied profile than the amount of free model parameters used to fit it with a synthetic profile (see, for example, Fig.~\ref{fig3}). Therefore we are able to obtain relatively good estimates of velocities ($V_{\rm r}$ and $V \sin{i}$) and abundances of all analysed elements that contribute to the studied line profile.

Using the second approach we can perform abundance analysis of spectra with V$\sin{i} >$ 40 km s$^{-1}$ that have a lot of blended line profiles. The spectral resolution and the S/N ratio of analysed spectra should be in this case high as well (R$\ge$60000 and S/N>250), because we fit each line profile using a higher amount of free model parameters (see Fig.~\ref{fig3}).

\section{Abundance analysis}
\label{analysis}

In our approaches for the abundance analysis we have used just the Stokes I spectra assuming a zero magnetic field (see Subsection~\ref{field}).
Simulation of each line profile was performed 4--5 times to reach the absolute minimum of $\chi^2$-function derived as the best fit result. Each new simulation was started with a different set of initial values of radial velocity, V$\sin{i}$ and abundance of chemical elements involved in the fitting routine. A post-simulation automatic procedure has been employed to compare the best fit data derived for each analysed profile in different simulations and select the best fit results with the smallest value of $\chi^2$-function.

\begin{figure*}
\begin{center}
\begin{tabular}{cc}
\includegraphics[width=2.1in,angle=-90]{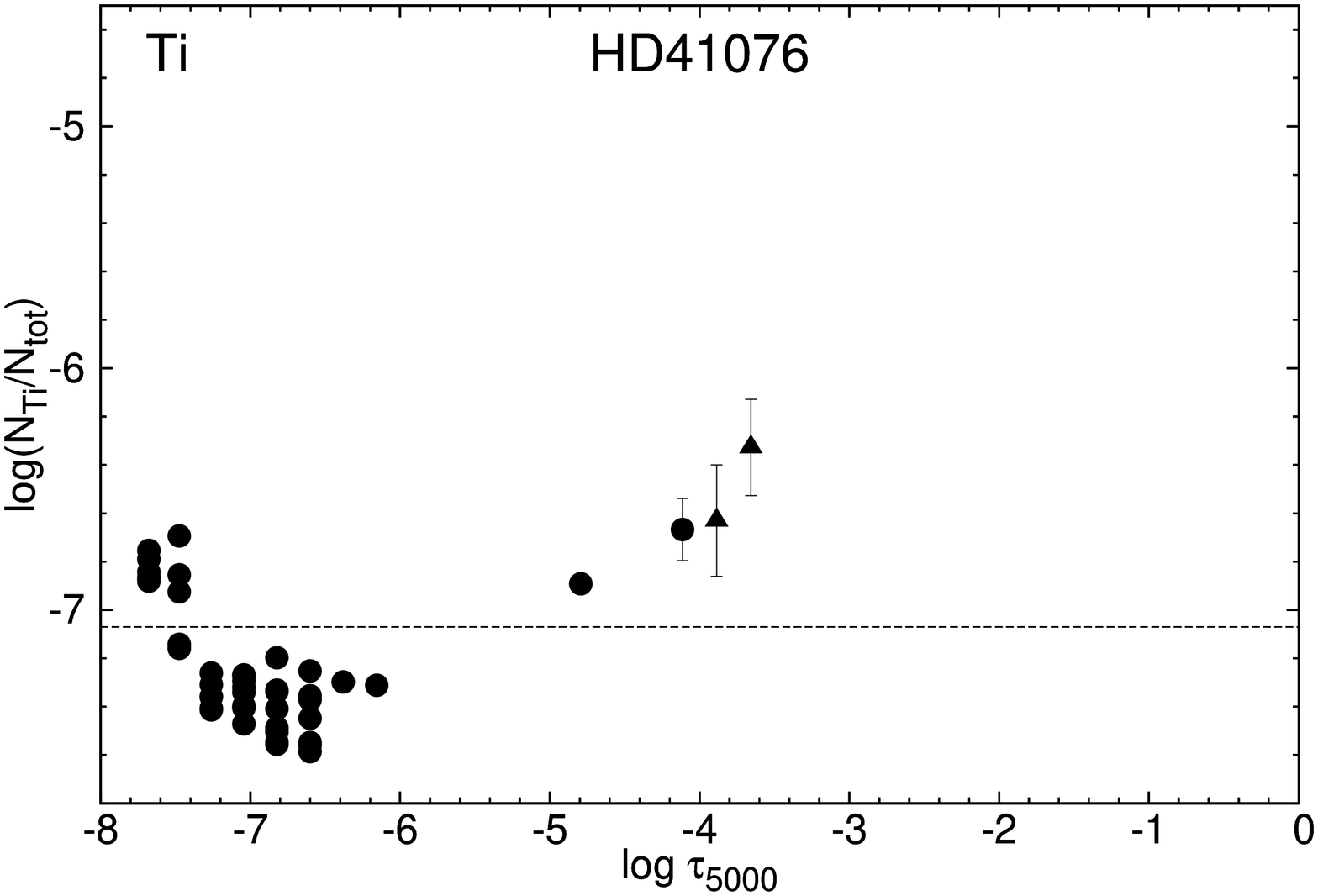} &
\includegraphics[width=2.1in,angle=-90]{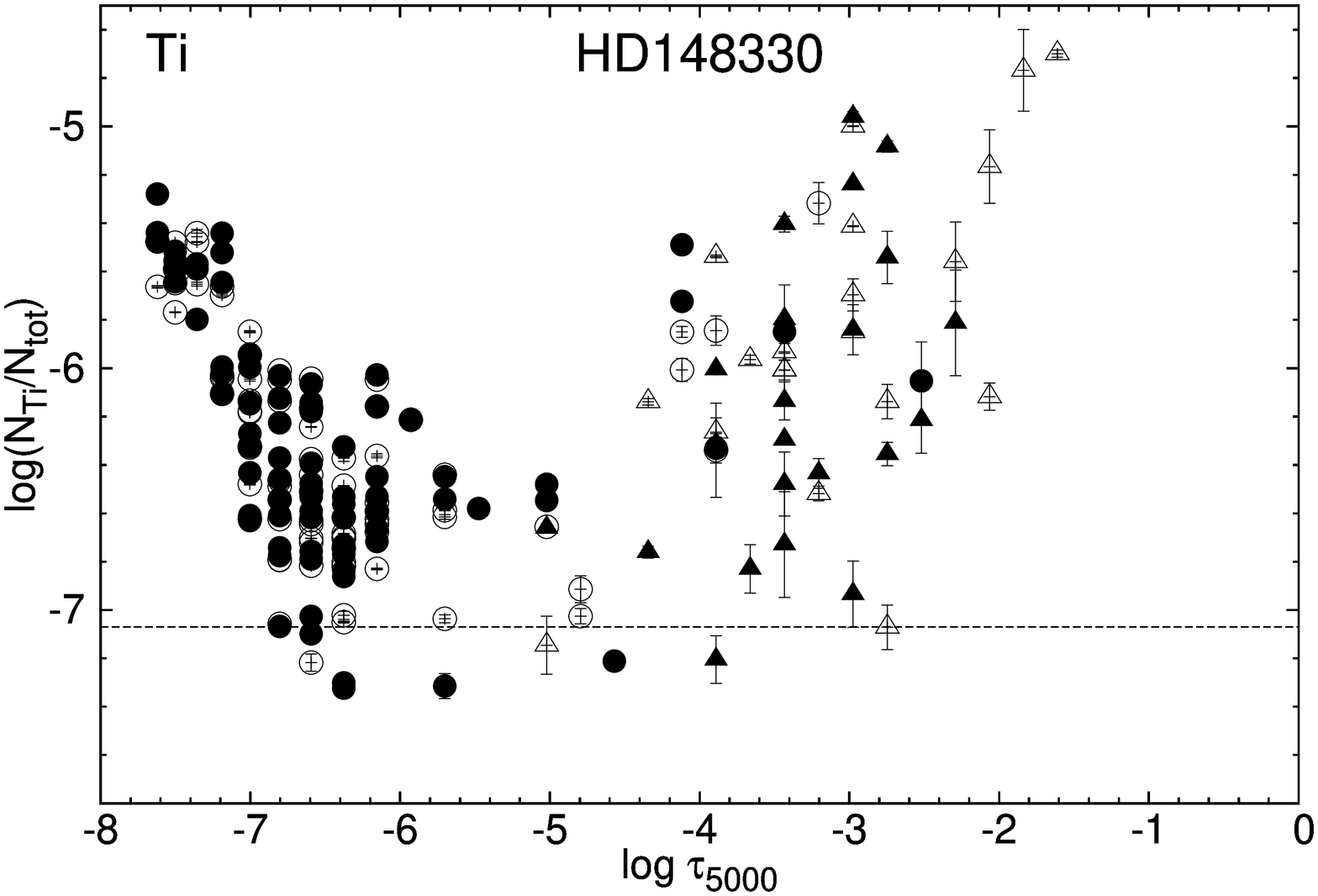}\\
\includegraphics[width=2.1in,angle=-90]{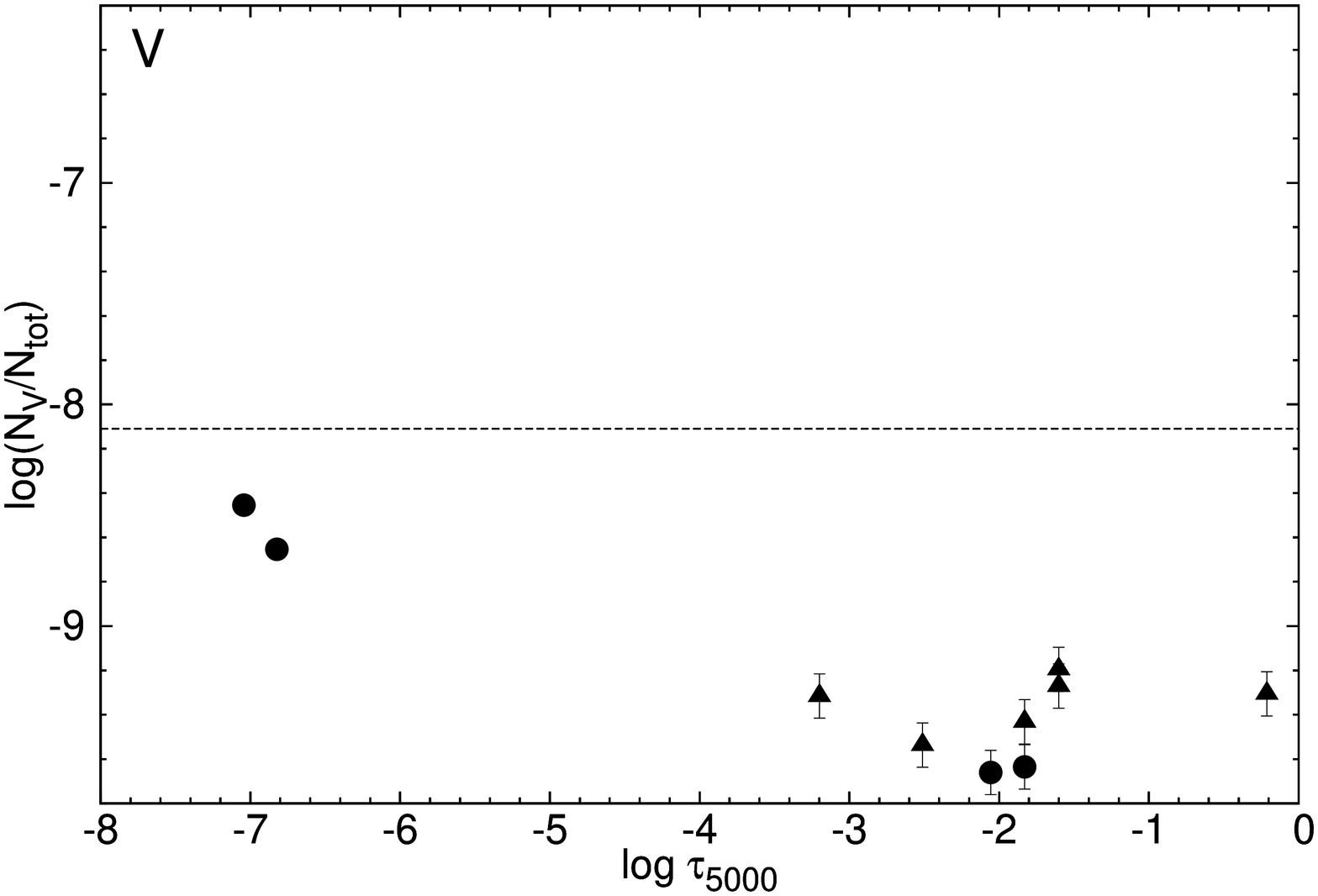} &
\includegraphics[width=2.1in,angle=-90]{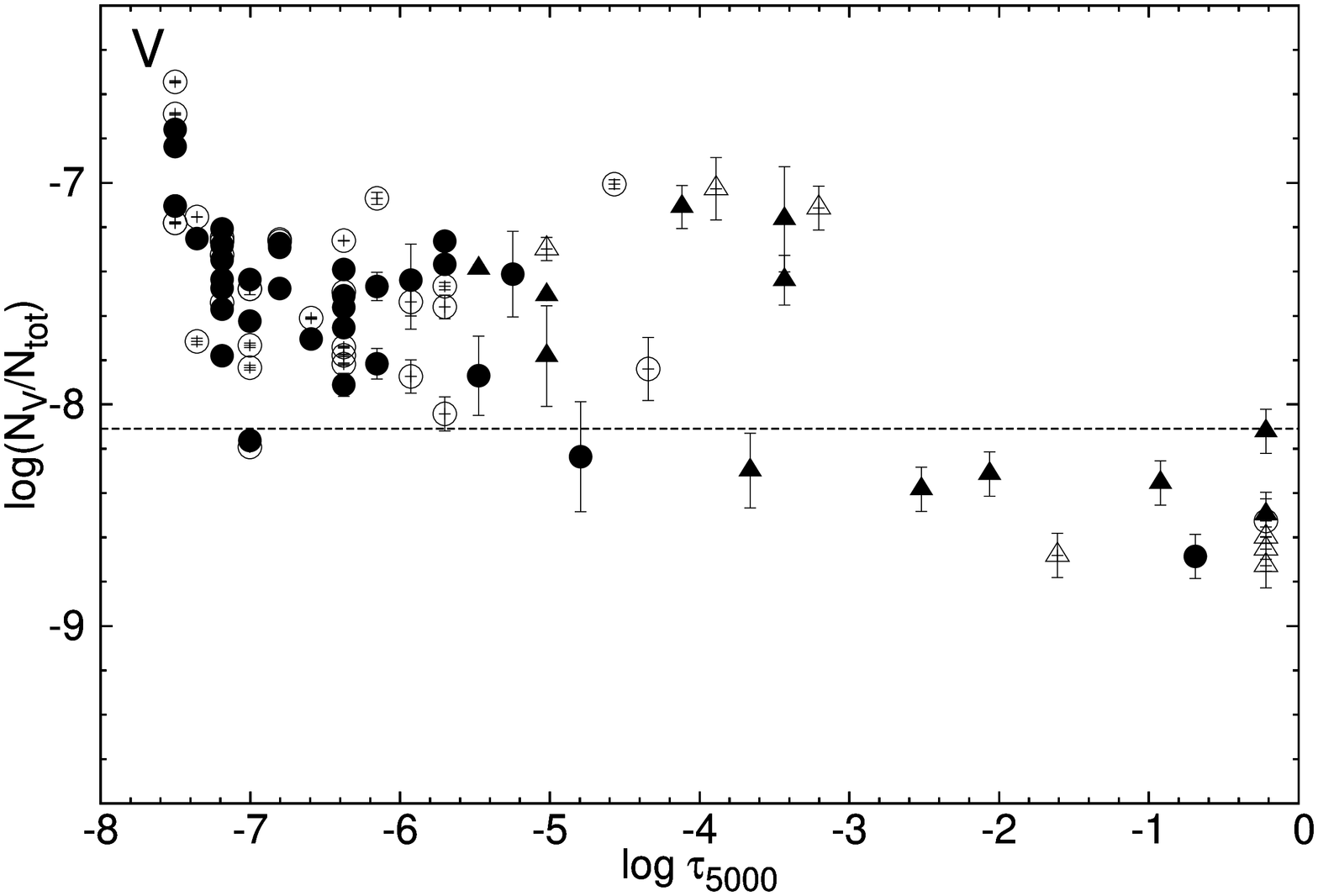}\\
\includegraphics[width=2.1in,angle=-90]{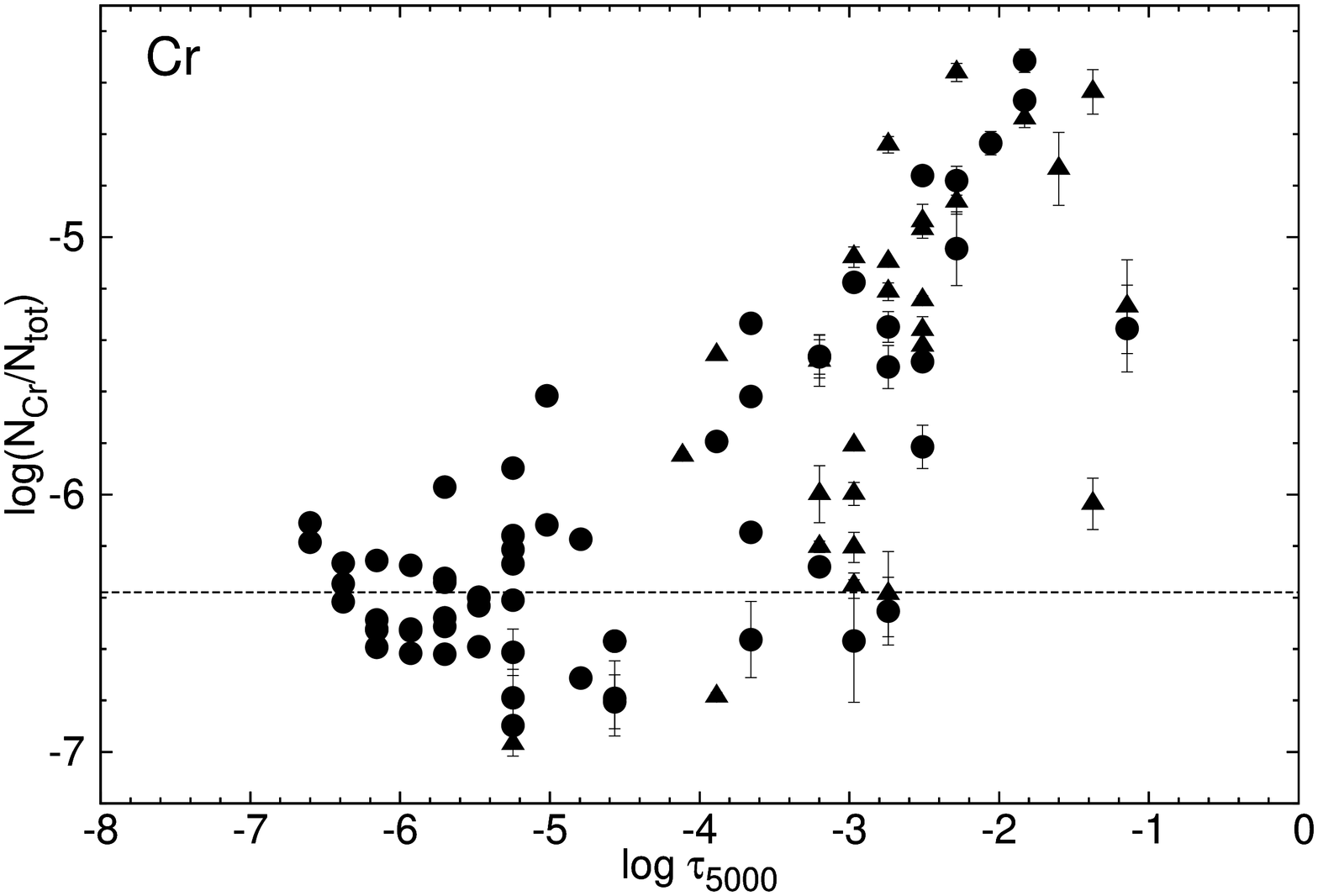} &
\includegraphics[width=2.1in,angle=-90]{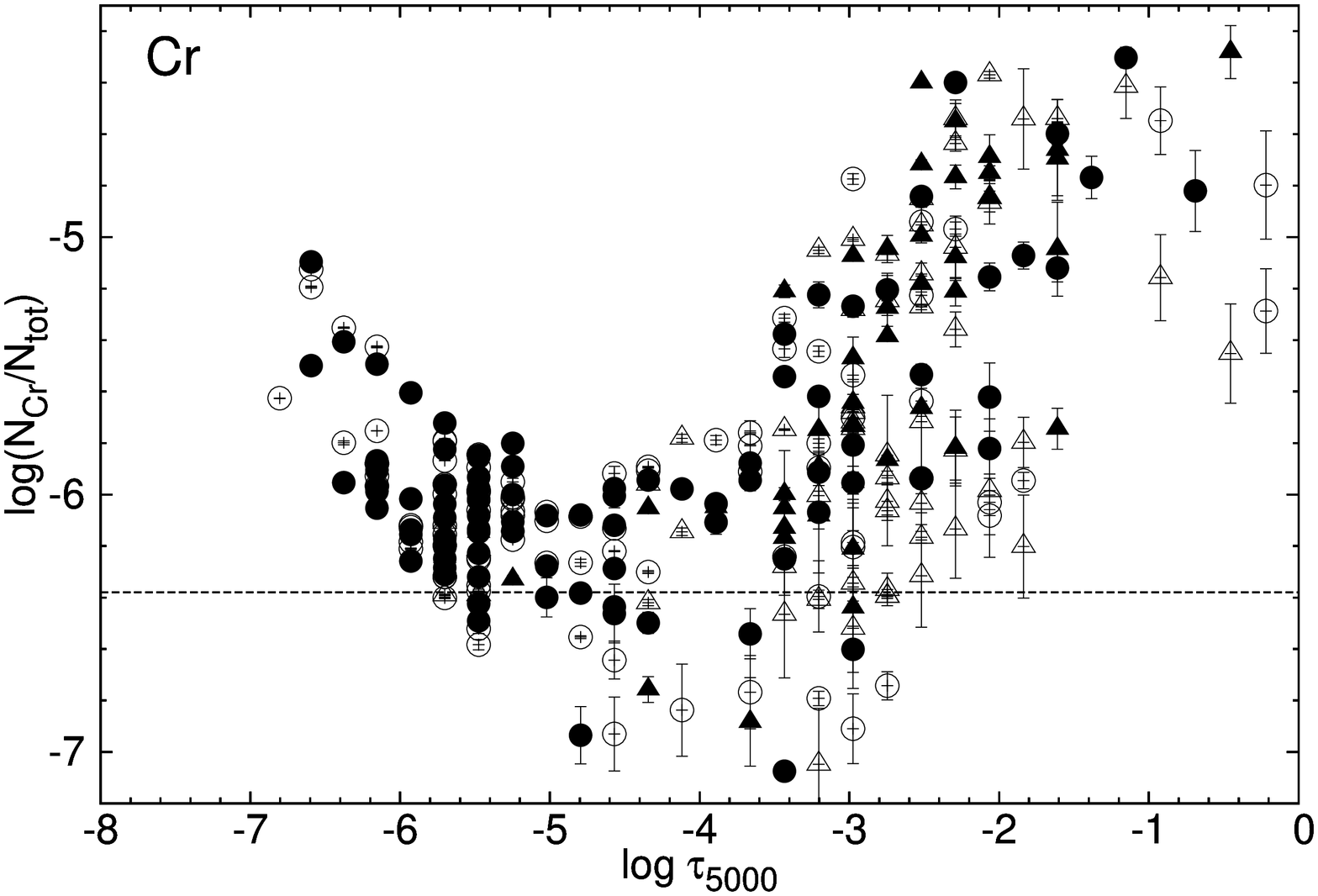}\\
\includegraphics[width=2.1in,angle=-90]{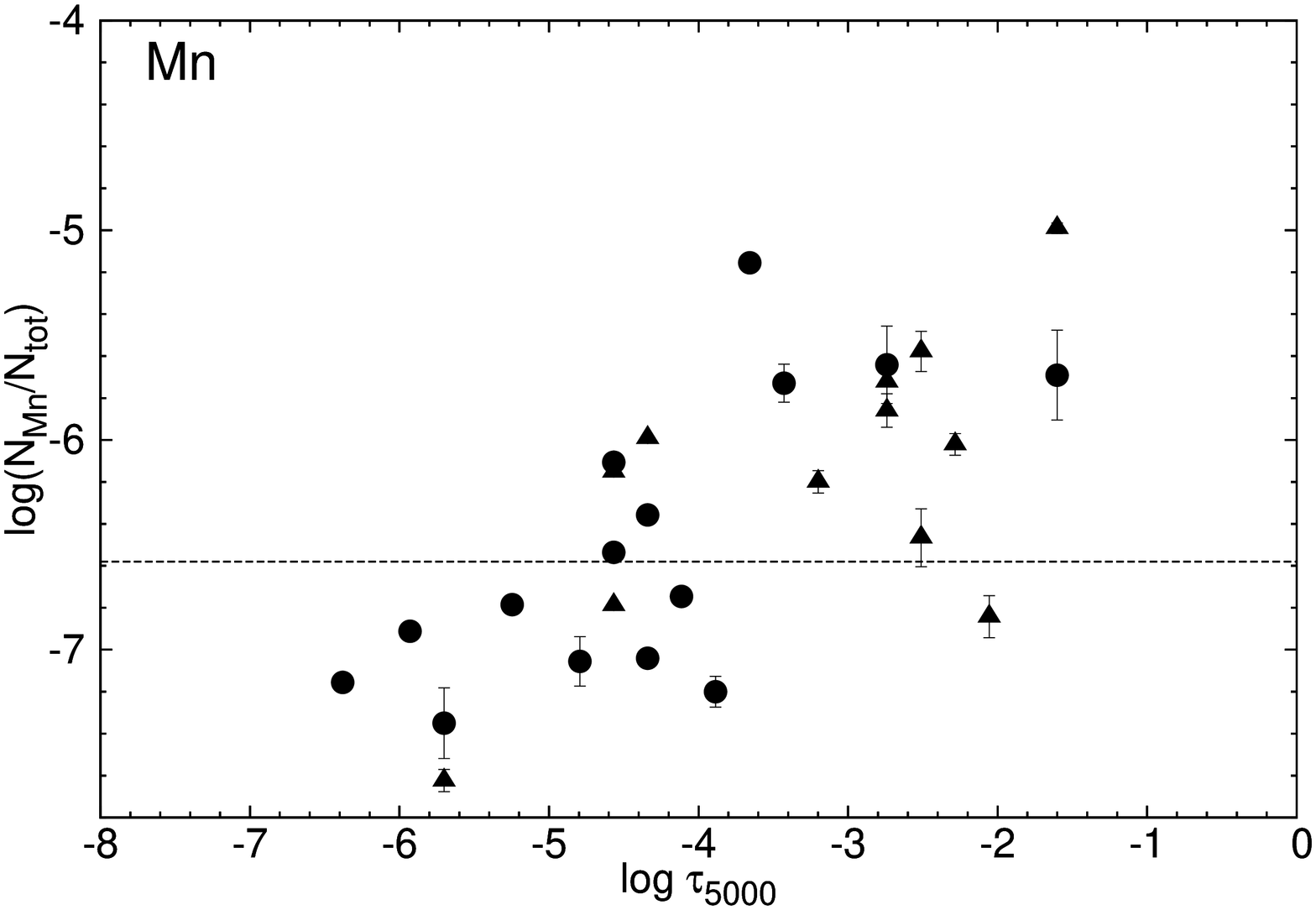} &
\includegraphics[width=2.1in,angle=-90]{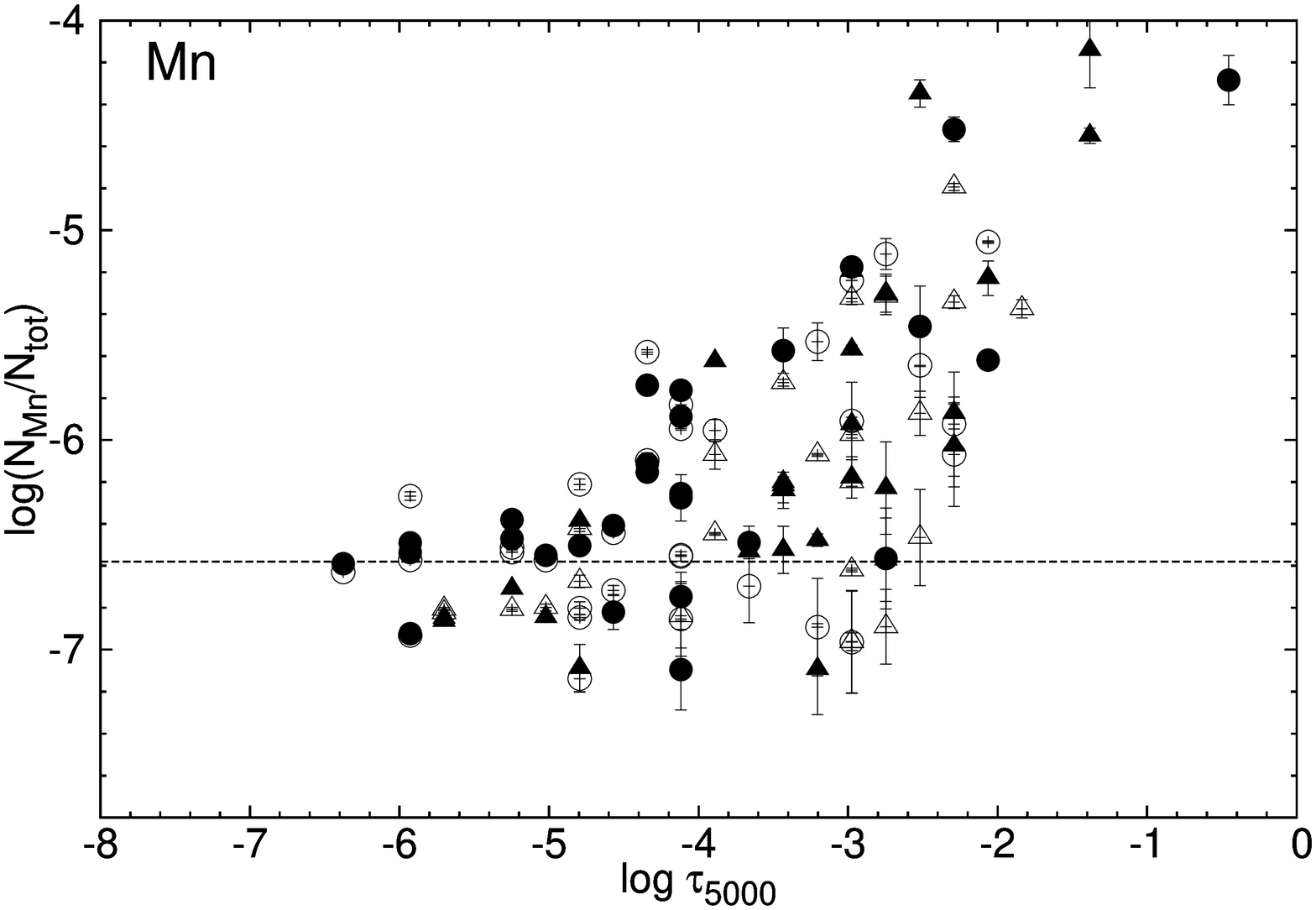} \\
\end{tabular}
\caption{ Distribution of titanium, vanadium, chromium and manganese abundance with optical depth respectively in HD~41076 (left panel) and HD~148330 (right panel). These data are obtained for the neutral (triangles) and once ionised (circles) ions of the aforementioned chemical elements using the second approach for abundance analysis (see Subsection~\ref{second}). In the case of HD~148330, the results derived for the first spectrum (observed on Feb.10, see Table~\ref{tab1}) are presented by filled symbols and those derived for the second spectrum (observed on Feb.15, see Table~\ref{tab1}) are shown by open symbols. Horizontal dashed line specifies the solar abundance for each studied element. }
\label{fig4}
\end{center}
\end{figure*}

Among the analysed line profiles, the same procedure has automatically selected for abundance analysis only those profiles that were fitted relatively well, i.e. for which the fitting routine resulted in a sufficiently small value of $\chi^2$-function \cite{Khalack+07}.
The upper cut-off limit of $\chi^2$-function depends on the S/N-ratio in studied spectrum (the higher S/N-ratio leads to a higher upper limit of $\chi^2$).
Profiles, for which we have obtained radial velocity and V$\sin{i}$ data that are significantly different from their average values, are also excluded from abundance analysis even if the derived value for $\chi^2$-function is relatively small (signifying a ``good" fit).

For each analysed ion whose lines contribute to a studied line profile, we determine its abundance and optical depth $\tau_{\rm 5000}$ that corresponds to a layer of core formation of the strongest line of this element in the spectral area of studied profile (see Subsection~\ref{second}). Abundance of ions responsible for formation of weak blends is usually determined with a low precision. Therefore for abundance analysis we accept only the abundance estimates with error bars smaller than 0.25 dex. Lists of the strongest spectral lines of the studied chemical elements, that contribute to the line profiles with a ``good" fit obtained implementing the second approach, are available online in the forme of Table~\ref{tab3} for HD~41076 and for the two spectra of HD~148330.

\begin{figure*}
\begin{center}
\begin{tabular}{cc}
\includegraphics[width=2.1in,angle=-90]{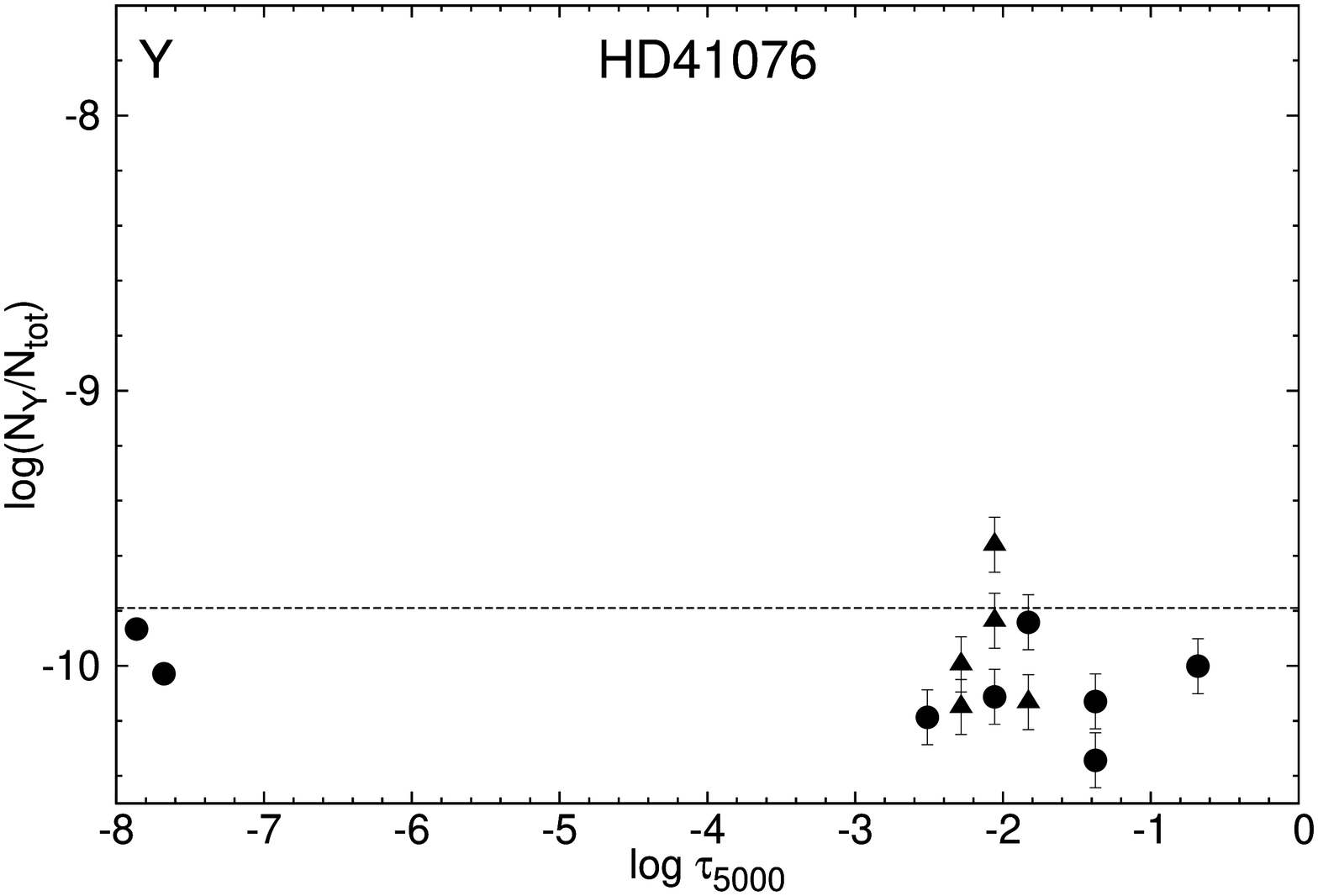} &
\includegraphics[width=2.1in,angle=-90]{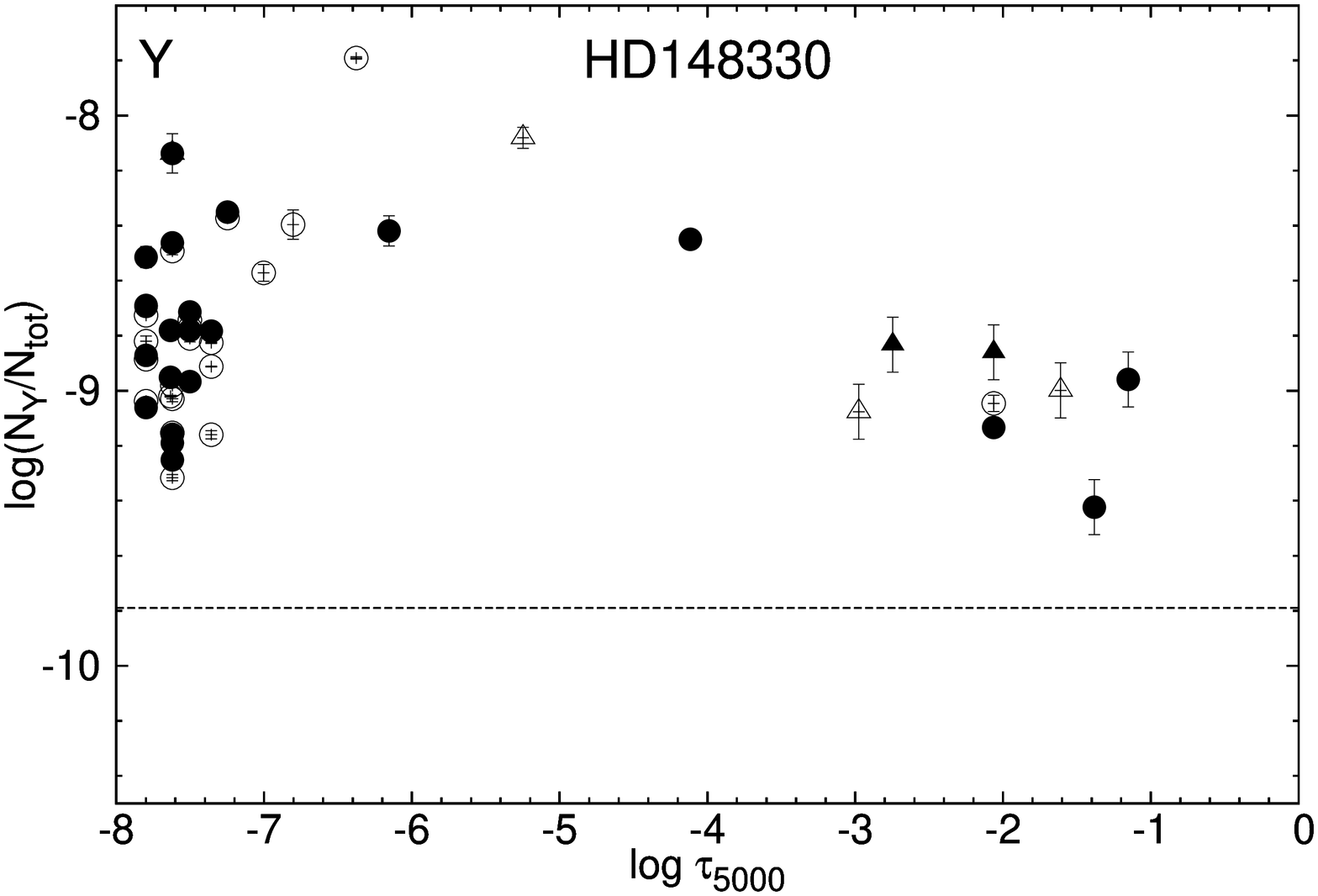}\\
\includegraphics[width=2.1in,angle=-90]{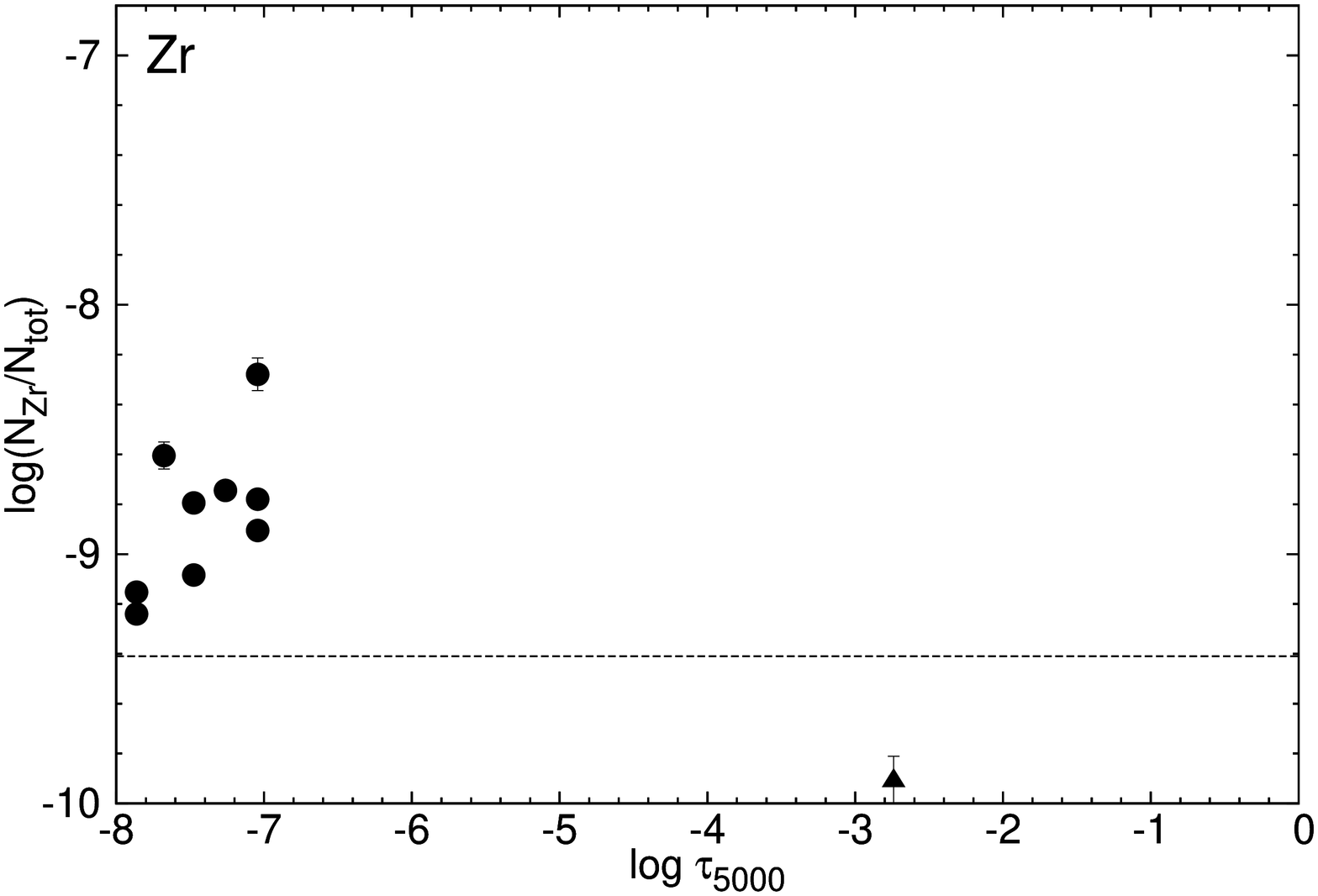} &
\includegraphics[width=2.1in,angle=-90]{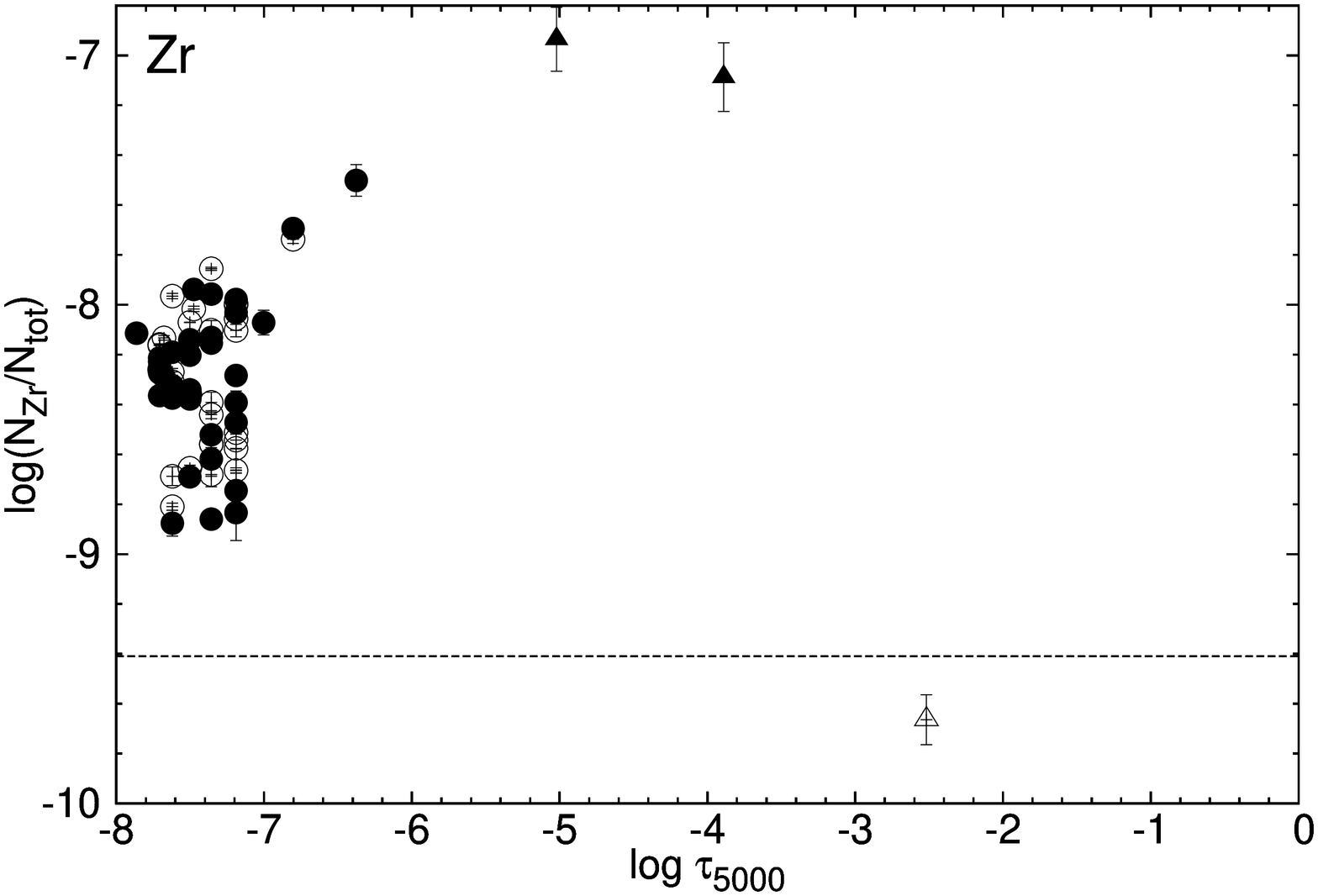}\\
\includegraphics[width=2.1in,angle=-90]{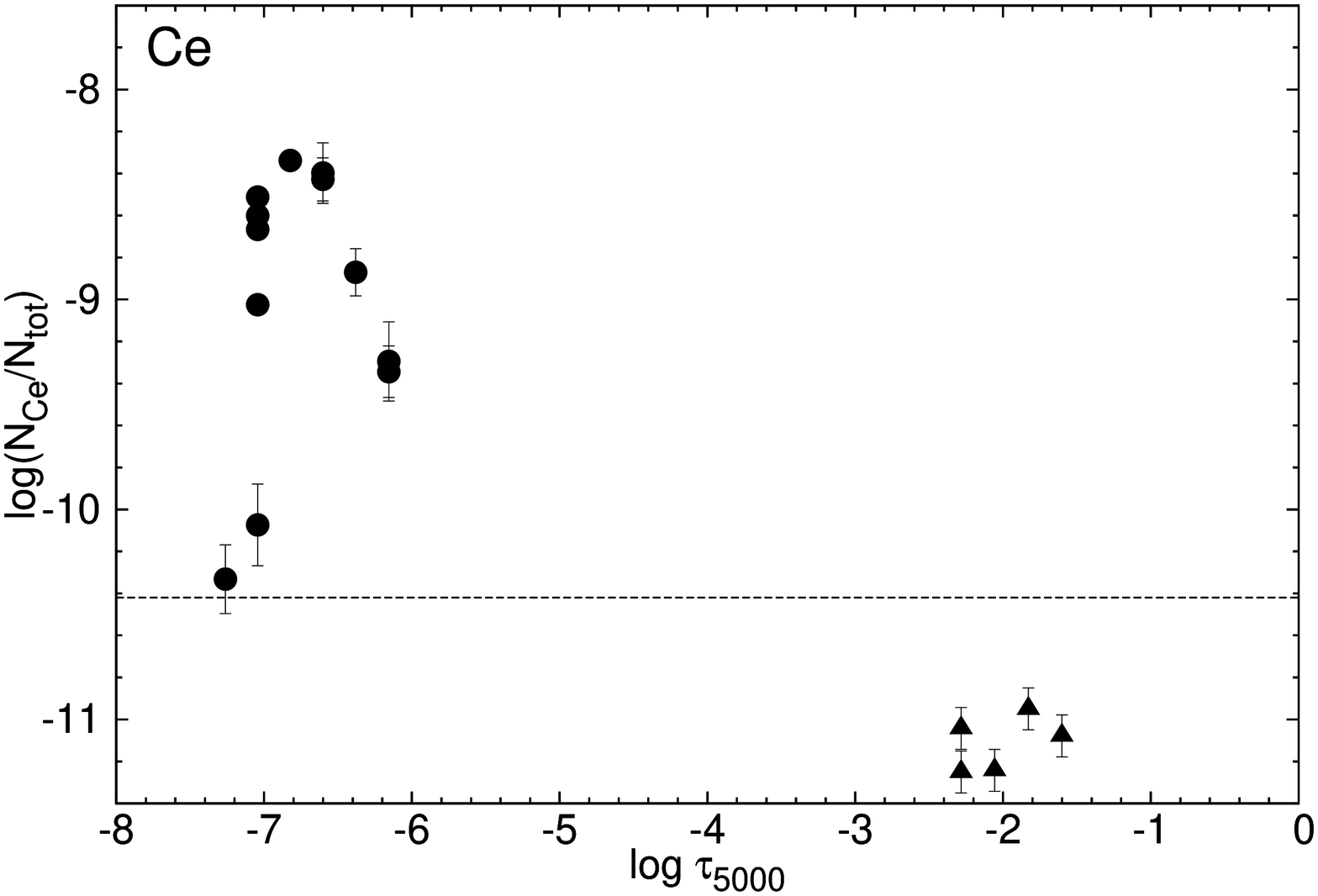} &
\includegraphics[width=2.1in,angle=-90]{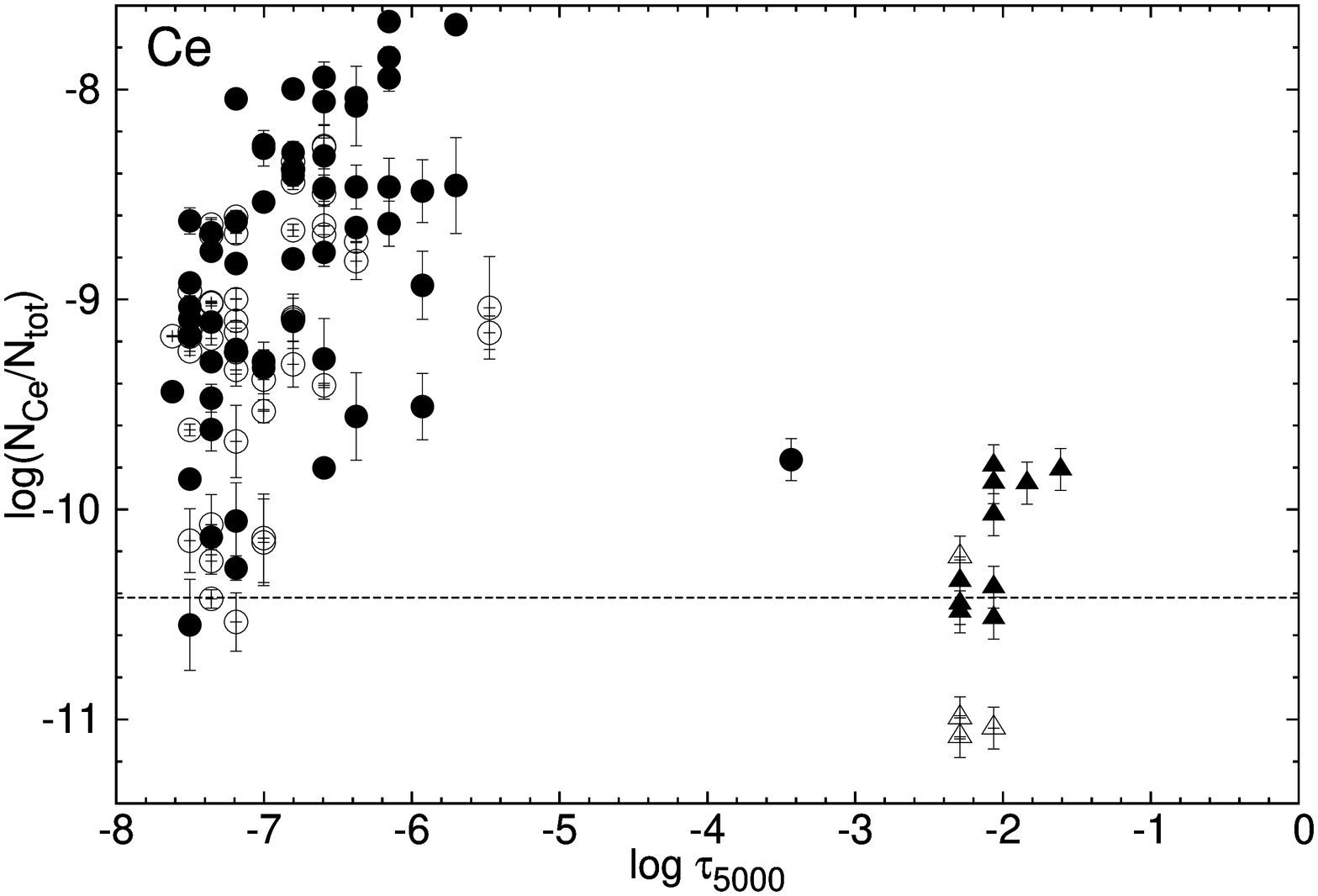}\\
\includegraphics[width=2.1in,angle=-90]{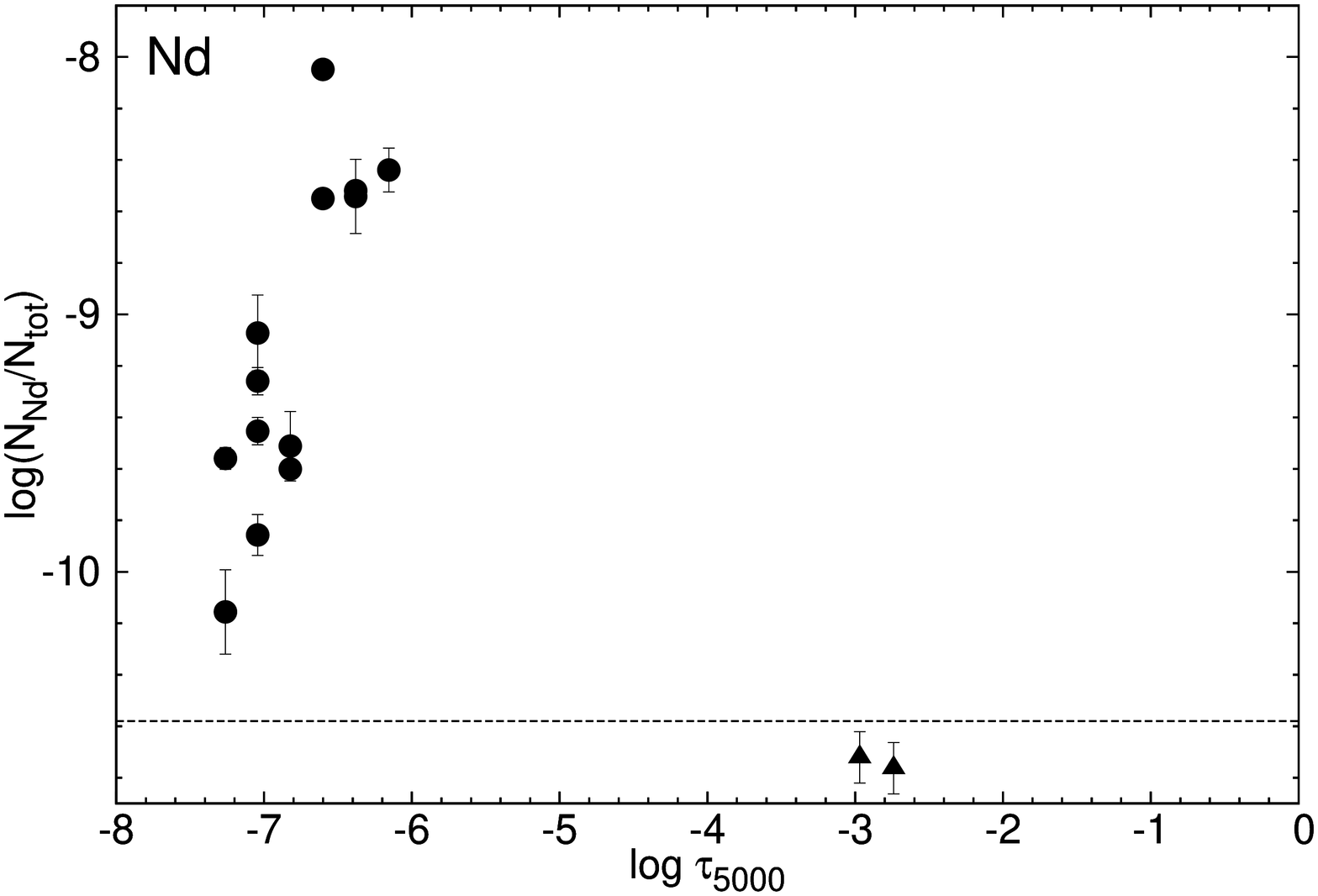} &
\includegraphics[width=2.1in,angle=-90]{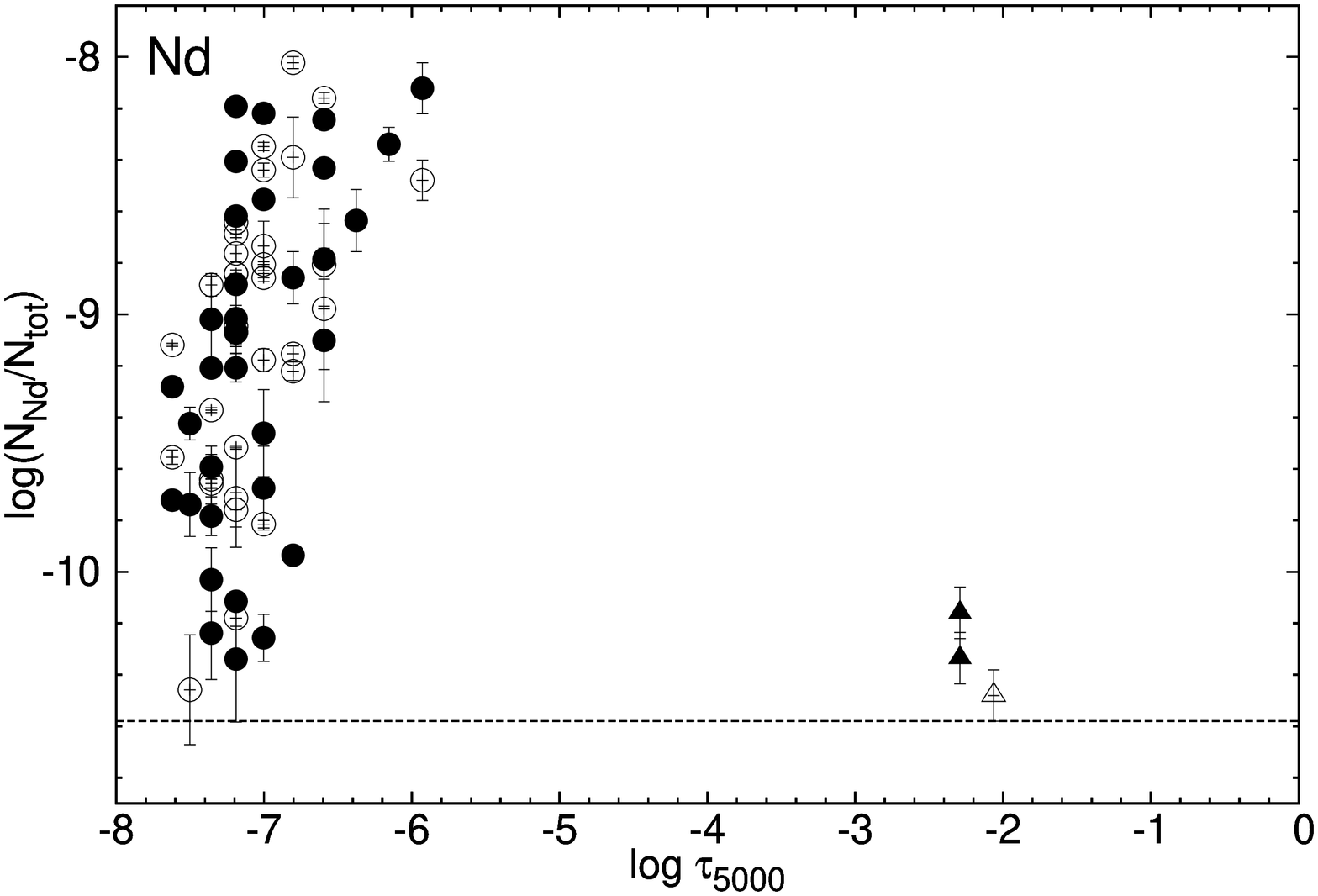}\\
\end{tabular}
\caption{ The same as at the Fig.~\ref{fig4}, but for the yttrium, zirconium, cerium and neodymium. }
\label{fig5}
\end{center}
\end{figure*}

\subsection{Average abundance estimates}
\label{results}

After selecting profiles with a ``good" fit, the aforementioned post-simulation automatic procedure has estimated an average abundance for each elements as well as for each ion of that element.
Tables~\ref{tab5},~\ref{tab5b} contain the average estimates of ions abundance derived from analysis of HD~41076 and HD~148330 spectra respectively. In Table~\ref{tab5}, the first column specifies the analysed chemical element. The 2nd and 3rd columns provide respectively an average abundance with estimation errors for "neutral" ions in stellar atmosphere of HD~41076 and a number of analysed profiles where the lines of neutral ions are the strongest ones (among the other spectral lines of the same element). The 4th and 5th, 6th and 7th columns provide the same information for once ionised ions and for both ions of this element in atmosphere of HD~41076 respectively. Two available spectra of HD~148330 (see Table~\ref{tab1}) have been analysed separately and the derived abundances for each chemical element are shown in two separate parts of the Table~\ref{tab5b}. The first seven rows are the same as in Table~\ref{tab5}, but provide information for the spectrum of HD~148330 observed on Feb. 10, while last six rows present similar data for the second spectrum of this star observed on Feb.~15.

Our analysis shows that in stellar atmosphere of HD~41076 the fluorine, vanadium, copper and gadolinium are in a significant deficit ($\sim$ -1.0 dex, see Tables~\ref{tab4},~\ref{tab5},~\ref{tab5b} and Fig.~\ref{fig4b}) with respect to their solar abundance \cite{Grevesse+10,Grevesse+15,Scott+15a,Scott+15b}, while molybdenum shows even stronger underabundance around -1.6 dex. Carbon, nitrogen, magnesium, calcium and samarium also show a significant underabundance ($\sim$ -0.3 -- -0.7 dex, see Table~\ref{tab5}) of neutral and once ionised ions in this star. It should be pointed out that using the first approach we found magnesium at its solar abundance (see Table~\ref{tab4}). The derived abundance of V\,{\sc i} appears to be smaller than the abundance of V\,{\sc ii}.
The detected Ti\,{\sc i} lines are mostly formed in the deeper atmospheric layers of HD~41076. Therefore, due to the vertical stratification of titanium abundance the Ti\,{\sc ii} is slightly underabundant, while the Ti\,{\sc i} appears to be significantly overabundant (see Fig.~\ref{fig4} and Table~\ref{tab5}).

Meanwhile, the chromium, cobalt, zirconium, barium and cerium show significant over-abundance ($\sim$ +0.4 -- +0.8 dex, see Table~\ref{tab5}) in the stellar atmosphere of HD~41076. At the same time due to vertical stratification of their abundance the neutral zirconium and cerium appear to be in deficit (see Fig.~\ref{fig5}), while the neutral chromium and cobalt show even higher abundance in the dipper atmospheric layers of this star. The Mn\,{\sc i} and Ni\,{\sc ii} are also significantly overabundant. The Sr\,{\sc ii}, Ce\,{\sc ii}, Nd\,{\sc ii} and Dy\,{\sc ii} show even stronger overabundance ($\sim$ +1.4 -- +2.8 dex) supporting an idea that HD~41076 is a chemically peculiar star (see Fig.~\ref{fig4b}).
Both approaches used for the abundance analysis result in similar abundances of the neutral and once ionised iron indicating that Fe\,{\sc ii} appears to be overabundant compared to Fe\,{\sc i}, which is slightly underabundant (see Fig.~\ref{fig2}).

Taking into account the error bars of the average abundance estimates, the helium, oxygen, neon, aluminium, silicon, phosphor, sulfur, argon, scandium, yttrium and praseodymium have solar abundance in HD~41076. For the sodium and zinc, we also have found that their abundance is close to the solar one. Nevertheless, this result is obtained from the analysis of only one spectral line for each element. Fig.~\ref{fig4b} presents the average abundance of studied chemical species with respect to their solar abundance and clearly proves that HD~41076 does show some abundance peculiarities as was indicated by Renson \& Manfroid \shortcite{Renson+Manfroid09}.

Analysis of the two available spectra of HD~148330 shows that the carbon, nitrogen, scandium and molybdenum are underabundant ($\sim$ -0.2 -- -0.4 dex, see Table~\ref{tab5b}) in stellar atmosphere of this star. The Ce\,{\sc i} seems to be significantly ($\sim$ -0.4 dex) underabundant, but the abundance of Ce\,{\sc ii}, whose line cores are mostly formed in the upper atmosphere, appears to be enhanced due the vertical stratification of cerium abundance (see Fig.~\ref{fig5}). The separate analysis of two spectra of HD~148330 has provided quite different estimates for the Ce\,{\sc i}, Ce\,{\sc ii} and Pr\,{\sc ii} average abundances (see Table~\ref{tab5b}). This also can be explained by the detected vertical and probably horizontal stratification of cerium abundance.

Meanwhile, the neutral and ionised ions of the sodium, phosphor, titanium, vanadium, chromium manganese and nickel are significantly overabundant ($\sim$ +0.4 -- +0.8 dex, see Table~\ref{tab5}), while the aluminium, cobalt, strontium, yttrium, zirconium, barium (Ba\,{\sc ii}), and rare earth elements (REE : Ce, Pr, Nd, Sm, Eu, Gd, Dy) show a strong ($\sim$ +1.0 -- +1.8 dex) overabundance in the stellar atmosphere of HD~148330. In this star, the strongest ($\sim$ +2.0 dex) overabundance is found for the Pr\,{\sc ii} (see Table~\ref{tab5b}). The Ne\,{\sc ii}, Si\,{\sc ii}, S\,{\sc ii}, Ca\,{\sc ii} and Fe\,{\sc ii} (see Fig.~\ref{fig2}) are slightly ($\sim$ +0.2 -- +0.4 dex) overabundant, while their neutral ions appear to have solar abundance in one spectrum or an enhanced abundance in the other spectrum (see Table~\ref{tab5b}).

The helium, oxygen, fluorine, magnesium, argon and copper show solar abundance in the stellar atmosphere of HD~148330 taking into account the error bars of their average abundance estimates.  Fig.~\ref{fig4b} shows a strong overabundance of REE and some metals in stelar atmosphere HD~148330 that confirms its classification by \v{Z}i\v{z}\v{n}ovsk\'{y} \shortcite{Ziga80} as a chemically peculiar star.

\subsection{Vertical abundance stratification}
\label{strat}

Fig.~\ref{fig2} shows the distributions of iron abundance with optical depths obtained employing the first (upper panel) and the second (lower panel) approaches for abundance analysis of iron in both stars (see Section~\ref{fit}). In the case of HD~148330, its two available spectra were combined for the abundance analysis employing the first approach. For each studied star, both methods result in similar change of iron abundance with optical depths, despite of different number of Fe\,{\sc i} and Fe\,{\sc ii} lines used for analysis.
The second approach is sensitive to the contribution of weak lines and therefore provides more data for the deeper atmospheric layers, which themselves have smaller precision of abundance estimate. In both stars the iron reaches its minimum abundance, which is close to its solar abundance, in the area of optical depths $\log\tau_{5000} \approx$ -4.0 -- -3.6 (see Fig.~\ref{fig2}). For both stars we can see that in this area of optical depths the Fe\,{\sc i} is statistically less abundant than the Fe\,{\sc ii}. Iron abundance tends to increase towards the upper atmosphere by +0.6 dex in HD~41076 and by +0.9 dex in HD~148330, and towards the deeper atmospheric layers by +0.9 dex and +1.2 dex in HD~41076 and HD~148330 respectively. In both stars the slopes of the increase of iron abundance with the optical depths are significant. Similar behaviour of iron abundance in the stellar atmospheres of HD~41076 and HD~148330 is not surprising because these stars possess similar effective temperature and gravity (see Table~\ref{tab2}).

The abundance change of titanium and chromium with optical depth in stellar atmosphere of HD~148330 are similar to the vertical stratification of the iron abundance. The titanium abundance reaches minimum corresponding to its solar abundance at $\log\tau_{5000} \approx$ -6.0 -- -4.5, while the chromium abundance reaches its minimum that is close to its solar abundance in the area of $\log\tau_{5000} \approx$ -5.0 -- -3.5 (see Fig.~\ref{fig4}). The titanium and chromium are represented approximately by 110 and 140 lines in the two available spectra of HD~148330 (see Table~\ref{tab5b}) and therefore we are able to detect a significant increase of their abundance towards the deeper and towards the upper atmospheric layers of this star. In the spectrum of HD~41076 the titanium and chromium are represented with much smaller amount of spectral lines (see Table~\ref{tab5}) and we can not see clearly the minima in their abundance change (see Fig.~\ref{fig4}). The chromium abundance tends to increase towards the deeper atmospheric layers of HD~41076 showing a significant slope.

In both stars the manganese also shows a tendency to increase its abundance towards the deeper atmosphere (see Fig.~\ref{fig4}). In the atmosphere of HD~41076 the vanadium is underabundant
and tends to decrease its abundance with the optical depths.
In the atmosphere of HD~148330 its abundance appears to be vertically stratified (overabundant in the upper atmosphere and underabundant in the deeper atmospheric layers). In the upper atmosphere of HD~148330 the vanadium abundance seems to reach some plateau ($\sim$ +0.6 dex above its solar abundance) in the area of optical depths of $\log\tau_{5000} \approx$ -7.0 -- -5.0 and after it tends to decrease towards the deeper atmospheric layers.

The abundance of yttrium seems to be vertically stratified in the atmosphere of HD~148330 showing a significant overabundance in the upper atmosphere and is less abundant at the deeper layers (see Fig.~\ref{fig5}). In the atmosphere of HD~41076 the yttrium is underabundant.

Abundance of the zirconium, cerium and neodymium shows a significant enrichment ($\gtrsim$ +0.5 dex above its solar abundance) in the upper layers of stellar atmosphere of both studied stars (see Fig.~\ref{fig5}). For HD~41076 we do not have enough data to search for vertical stratification of Zr, Ce and Nd abundance.
In the upper atmosphere of HD~148330 the abundance of neodymium seems to increase from the solar level at $\log\tau_{5000} \approx$ -7.0 by 2.0 dex at $\log\tau_{5000} \approx$ -6.0. We have found a similar tendency of increase of neodymium abundance from the analysis of two available spectra of HD~148330, but the area of optical depth where we observe the abundance increase is quite narrow (see Fig.~\ref{fig5}). The abundance of Zr, Ce, Sm and Gd shows similar behaviour to the neodymium abundance in the upper atmosphere of HD~148330.
For praseodymium we do not have have enough data to find out if its abundance is vertically stratified or not. Meanwhile, zirconium abundance seems to increase in the area of $\log\tau_{5000} \approx$ -7.5 -- -5.0. The abundance of cerium first tends to increase towards the deeper atmospheric layers and after drops down to the level of its solar abundance (see Fig.~\ref{fig5}).

\begin{table}
\begin{center}
\caption{Fundamental stellar parameters employed in the second approach and the velocities obtained by averaging the data from all (total number is given in the brackets) analysed line profiles.}
\label{tab7}
\begin{tabular}{lccc}\hline
Parameter & HD~41076 & \multicolumn{2}{c|}{HD~148330}\\
 & & Feb. 10$^a$ & Feb. 15$^a$ \\
\hline
$T_{\rm eff}$ (K)         & 9483$\pm$100  &  \multicolumn{2}{c|}{9303$\pm$100}   \\
$\log{g}$                 &  3.69$\pm$0.10 &  \multicolumn{2}{c|}{3.70$\pm$0.10}   \\
${\rm [M/H]}$             &  -0.5$\pm$0.1 &   \multicolumn{2}{c|}{0.0$\pm$0.1} \\
$V \sin{i}$ (km s$^{-1}$) &  7.4$\pm$0.2 (350)  &   9.5$\pm$0.2 (580) & 9.5$\pm$0.2 (555)\\
$V_{\rm r}$ (km s$^{-1}$) & -13.7$\pm$0.2 (350) &  -3.5$\pm$0.2 (580) & -3.5$\pm$0.2 (555)\\
\hline
\end{tabular}
\end{center}
{\it Notes:} $^a$results for two separate spectra (see Table~\ref{tab1}) 
\end{table}

\section{Discussion}
\label{discus}

Vertical stratification of element abundances in a hydrodynamically stable atmosphere can be studied through the analysis of multiple line profiles that belong to the same ion of the investigated element, for instance, using the modified version of ZEEMAN2 code \cite{Khalack+Wade06,Khalack+07}.
This approach has been successfully employed to search for the signatures of vertical abundance stratification in the atmospheres of several BHB stars \cite{Khalack+08,Khalack+10,LeBlanc+10}, one post-HB star \cite{Khalack+14}, and several HgMn stars \cite{Thiam+10}. A number of the upper main sequence CP stars observed in the frame of the VeSElkA project \cite{Khalack+LeBlanc15a,Khalack+LeBlanc15b} have been studied using the aforementioned approach. The abundance analysis resulted in detection of indications of vertical stratification of iron abundance in stellar atmosphere of HD~95608 and HD~116235 \cite{Khalack+13,LeBlanc+15} and of chromium abundance in stellar atmosphere of HD~22920 \cite{Khalack+Poitras15}.

Here, we presented our estimate of the effective temperature, gravity and metallicity (see Fig.~\ref{fig1} and Table~\ref{tab2}) in chemically peculiar stars HD~41076 and HD~148330, their radial velocities, and $V \sin{i}$ obtained from the analysis of several hundreds of spectral line profiles selected from the available spectra of each star (see Table~\ref{tab7}).
Fitting of the Balmer line profiles with the help of FITSB2 code \cite{Napiwotzki+04} provides similar values of $T_{\rm eff}$ and $\log{g}$ for both studied stars (see Table~\ref{tab2}).
For HD~41076 and HD~148330 the obtained values of $T_{\rm eff}$ are consistent with the results derived from the $c_{\rm 1}$ \cite{Napiwotzky+93} and $\rm (B-V)_{0}$ \cite{Netopil+08} photometric temperature calibrations (see Tables~\ref{tab2},~\ref{tab3b}).
The fitting procedure has used the synthetic fluxes calculated for different metallicities \cite{Husser+13,Khalack+LeBlanc15a} and resulted in the solar value of metallicity for HD~148330 and in M=-0.5 for stellar atmosphere of HD~41076.
The interpolation of data employing the izochrones \cite{Bressan+12} closest to the position of HD~41076 and HD~148330 in the HR diagram (see Fig.~\ref{fig1a}) allowed us to derive their age, mass and surface gravity. For both stars the derived gravities are consistent with the values of $\log{g}$ obtained from the fitting of Balmer line profiles taking into account the estimation errors (see Tables~\ref{tab2},~\ref{tab3b}).

The fitting of the Balmer line profiles with the FITSB2 code and the fitting 
of line profiles that belong to other chemical elements employing the first \cite{Khalack+Wade06,Khalack+07} and the second (see Subsection~\ref{second}) approaches for abundance analysis provide consistent values of radial velocity for both studied stars (see Tables~\ref{tab2}  and \ref{tab7}). These data are in a good accordance with the previously published data $V_{\rm r}$ = -12.9$\pm$ 2.8 km s$^{-1}$ for HD~41076 and $V_{\rm r}$ = -3.8$\pm$ 1.7 km s$^{-1}$ for HD~148330 \cite{Gontcharov2006}, and $V_{\rm r}$ = -4.0$\pm$ 0.9 km s$^{-1}$ for HD~148330 \cite{Ziga+Romanyuk90} taking into account the estimation errors. Meanwhile, our estimates of the rotational velocity $V \sin{i}$ (see Tables~\ref{tab4} and \ref{tab7}) appear to be much smaller than the values ($V \sin{i}$ = 14 km s$^{-1}$ for HD~41076 and $V \sin{i}$ = 18 km s$^{-1}$ for HD~148330) found by Royer et al. \shortcite{Royer+02} for both stars. Nevertheless, in the case of HD~148330 our result $V \sin{i}$ = 9.6$\pm$0.1 km s$^{-1}$ is close to the values 11.6 km s$^{-1}$ measured by \v{Z}i\v{z}\v{n}ovsk\'{y} \& Romanyuk \shortcite{Ziga+Romanyuk90}, 10 km s$^{-1}$ derived by Abt \& Morell \shortcite{Abt+Morell95}, and 10.5$\pm$0.5 km s$^{-1}$ obtained by Shorlin et al. \shortcite{Shorlin+02} for its rotational velocity. Our estimates of the rotational velocities are obtained from the analysis of a considerable number of line profiles (see Table~\ref{tab7}) with a significantly higher signal-to-noise ratio and spectral resolution (see Section~\ref{obs} and Table~\ref{tab1}) than those used by Royer et al. \shortcite{Royer+02}. Therefore, we consider our $V \sin{i}$ values obtained for both studied stars as more reliable and accurate data.

Comparative analysis of the results obtained with the first \cite{Khalack+Wade06,Khalack+07} and the second (see Subsection~\ref{second}) approaches to abundance analysis clearly shows the advantage of the second one. The second approach allows to prepare in a semi-automatic mode the input data for the modified ZEEMAN2 code for each line profile and to analyse with it several hundreds of those profiles in sequence during a single run of the code. To reach better fit (smaller value for the $\chi^2$-function) we need to repeat the simulation a few times imposing different initial values of the searched parameters. Both approaches provide very similar results of abundance analysis for all studied chemical elements (see Tables~\ref{tab4},~\ref{tab5} and \ref{tab5b}), but the second approach is semi-automated and results in much bigger amount of abundance estimates spread over the whole area of analysed optical depths. For each chemical element, more information can be derived for the deeper atmospheric layers trough the estimation of contribution of blends to the analysed profiles. Respectively the error-bars for abundance estimates at the deeper atmospheric layers are much bigger that the error-bars derived for the upper atmosphere (see Figs.~\ref{fig4},~\ref{fig5}). The data derived for the deeper atmosphere has also a higher distribution on abundance scale. That can be explained in part by the errors in oscillator strengths and dumping coefficients of weak spectral lines found in the NIST \cite{Kramida+15} and VALD3 \cite{Kupka+00,Ryab+15} databases.

Our estimates of average abundance show that the carbon, nitrogen and molybdenum are in deficit, while the chromium, cobalt, nickel, strontium, zirconium, barium, cerium, neodymium and dysprosium appear to be overabundant in both studied stars (see Tables~\ref{tab4},~\ref{tab5},~\ref{tab5b}). The abundance of iron  and manganese seems to be slightly enhanced in HD~41076 and HD~148330. The abundances of Mg, Ca, Sc, V, Y and Sm in the stellar atmosphere of HD~41076 are significantly lower than their solar abundances, while the F, Cu and Gd are in strong deficit ($\sim$ -0.8 -- -1.0 dex). In the stellar atmosphere of HD~148330 the Co, Sr, Y, Zr, Ba and REE are strongly enhanced showing overabundance $\sim$ +1.0 -- +2.0 dex. The other metals (like Ne, Na, Al, Si, P, S, Ca, Ti, V) are also overabundant in this star.
Our estimates of the Ca, Cr, Fe, Ni, Y, abundance from the analysis of HD~148330 spectra are in good accordance with the results derived by \v{Z}i\v{z}\v{n}ovsk\'{y} \shortcite{Ziga80} for this star. The abundance of Sc, Ti, V, Ce, Eu, Gd obtained in this study are smaller than the abundance found for these elements by the aforementioned author. Meanwhile, our estimates of the Si, Sr and Zr abundance appear to be slightly higher than the ones reported by \v{Z}i\v{z}\v{n}ovsk\'{y} \shortcite{Ziga80} for HD~148330.

The peculiarities of chemical abundance found in this study for HD~148330 and HD~41076 argue that these objects belong to the class of CP stars.
Both stars show an abundance peculiarity for different chemical species (see Fig.~\ref{fig4b}) similar to those found for other CP stars by Erspamer \& North \shortcite{Erspamer+North03} from the analysis of ELODIE spectra. Taking into account that HD~148330 has been classified previously as a SiSr star \cite{Cowley+69,Renson+Manfroid09} we confirm the strong enhancement of strontium
abundance and the overabundance of Si\,{\sc ii} in the stellar atmosphere of HD~148330 (see Table~\ref{tab5b}).
Apparently, the iron peak elements and the REE are in general much more abundant in the stellar atmosphere of HD~148330 than in HD~41076, but some metals and REE are in deficit in HD~41076. This fact may suggest a different chemical evolution of HD~148330 and HD~41076, that otherwise posses similar fundamental stellar parameters (see Table~\ref{tab3b}) except for their luminosity and age.

Use of the two approaches to study a variation of element's abundance with optical depth provides the same results for vertical stratification of iron abundance for each studied star (see Fig.~\ref{fig2}). For other chemical species (like Si, Ti, Cr, etc.) we also have found similar variation of their abundance with the optical depth using the two applied approaches.
In stellar atmospheres of HD~41076 and HD~148330 iron abundance reaches its minimum at the area of optical depths $\log\tau_{5000} \approx$ -4.0 -- -3.6, where the Fe\,{\sc i} is statistically less abundant than the Fe\,{\sc ii}. The iron abundance tends to increase by +0.9 dex (and more) towards the upper and deeper atmospheric layers with respect to this area in both stars. We also have found similar behaviour for the titanium and chromium abundances in the stellar atmosphere of HD~148330, where these elements are represented with a significant number of lines (see Fig.~\ref{fig4}). We found that the abundance of titanium has a tendency to increase towards the upper atmosphere, while the abundance of chromium increases towards the deeper layers in the stellar atmosphere of HD~41076. Manganese also appears to be vertically stratified in both studied stars and its abundance increases towards the deeper atmospheric layers.
Vanadium tends to decrease its abundance with optical depth in both stars. Nevertheless, in the upper atmospheric layers of HD~148330 vanadium appears to be overabundant, while in HD~41076 it is underabundant.

For both stars we have detected a significant enhancement of some REE (Ce, Nd and Dy in HD~41076, and Ce, Pr, Nd, Sm, Eu, Gd and Dy in HD~148330) in their upper atmospheric layers. Zirconium is also significantly overabundant in the upper atmosphere of both stars, while a strong yttrium enhancement we have found only in the upper atmospheric layers of HD~148330 (see Fig.~\ref{fig5}).
In this star abundance of yttrium and cerium first increases with optical depth and after drops down almost to its solar value. These results confirm the known fact first reported by Mashonkina et al. \shortcite{Mashonkina+05} that some CP stars show an overabundance of REE in their upper atmospheric layers.

\v{Z}i\v{z}\v{n}ovsk\'{y} \& Romanyuk \shortcite{Ziga+Romanyuk90} have found for HD~148330 a variable mean longitudinal magnetic field that can be represented by the inclined dipole model. For both stars studied here we have estimated their mean longitudinal magnetic field $<B_{\rm z}>$ (see Table~\ref{tab1}) from analysis of $H_{\rm \alpha}$ line core using the method described by Landstreet et al. \shortcite{Landstreet+15}. Our estimates of $<B_{\rm z}>$ are consistent with the previously published results \cite{Ziga+Romanyuk90,Shorlin+02} and show that the mean longitudinal field is very small ($<|B_{\rm z}|> \sim$ 100G). The magnetic field measurements in both stars are statistically very close to zero taking into account the measurement errors and therefore we have not included it into the procedures of line profile fitting and abundance analysis (see Subsection~\ref{field}). Such a small field can only affect a formation of spot structure with an overabundance of some metals or REE very close to magnetic poles \cite{Alecian+Stift10}. Analysis of the two available spectra of HD~148330 does not show a significant line profile variability that may indicate a presence of some spot structure. Taking into account that the derived mean longitudinal magnetic field is close to zero a possible explanation here can be that we do not directly see the magnetic poles in the case of these two spectra.

Nevertheless, in a hydrodynamically stable atmosphere even with a small or no magnetic field the atomic diffusion can lead to a vertical stratification of elements abundance \cite{Michaud+15}.
We would like to point out that the behaviour of the vertical stratification of iron abundance found in both studied stars is similar to the one derived by Alecian \& Stift \shortcite{Alecian+Stift10} from theoretical simulations for the zero-field equilibrium stratification of iron (see their Fig.~8), but in a stellar atmosphere with $T_{\rm eff}$ =12000K and $\log(g)$=4.0. Respectively, our data on vertical stratification of elements abundance can be used to improve the modeling of element stratifications and time-dependent atomic diffusion in magnetic ApBp stars and in non-magnetic CP stars as well \cite{Alecian+Stift10,Stift+Alecian12,Stift+Alecian16}, and for development and further improvement of realistic stellar atmosphere models \cite{HBH+00,LeBlanc+09}.

Both approaches can be used to analyse spectra of magnetic CP stars if the signal in Stokes V (Q or U) profile of an individual line (observed for different rotational phases) is strong enough to derive information about the configuration of magnetic field. In this case the characteristics of magnetic field structure are treated as free model parameters and the field strength at the magnetic pole should be at least few kG \cite{Khalack+Landstreet12}. The second approach is more flexible because it can deal with a blended line profile enlarged by the magnetic broadening, although here we have to take into account maps of horizontal abundance stratification of chemical species that contribute to this profile \cite{Khalack+Wade06}.

For HD~41076 and HD~148330 we have calculated the models of stellar atmosphere assuming a homogeneous abundance distribution of chemical elements. The detected vertical abundance stratification of the aforementioned metals can affect the structure of stellar atmosphere models used for the abundance analysis. Therefore, as the next step we plan to calculate empirical model atmospheres for these stars similar to those developed by Shulyak et al. \shortcite{Shulyak+09} employing the LLmodels code \cite{Shulyak+04}, and use them to refine our results of abundance analysis. The empirical models of stellar atmosphere could be simulated with a new modified version of PHOENIX 16 code. We expect that those models will take into account detailed (with respect to the optical depths) stratification profiles derived in the present study for each studied chemical element.

\section*{Acknowledgments}

The authors are grateful to the anonymous referee for his (her) valuable suggestions that allowed to improve this manuscript and to Dr. Schlafly for the help with determination of interstellar reddening to the studied stars.
G. Gallant and C. Thibeault are sincerely thankful to the Facult\'{e} des \'{E}tudes Sup\'{e}rieures et de la Recherch de l'Universit\'{e} de Moncton for the financial support of their research.
G. Gallant is also grateful to ``Le fond en astronomie de l'Universit\'{e} de Moncton" for the financial assistance to carry out the abundance analysis of HD~41076 using the first approach.
Part of calculations have been done on the supercomputer {\it briarree} of the University of Montreal, under the guidance of Calcul Qu\'{e}bec and Calcul Canada. The use of this supercomputer is funded by the Canadian Foundation for Innovation (CFI), NanoQu\'{e}bec, RMGA and Research Fund of Qu\'{e}bec - Nature and Technology (FRQNT).
This paper has been typeset from a \TeX/\LaTeX\, file prepared by the authors.

\label{lastpage}

\end{document}